\documentclass[onecolumn,11pt,oneside,draftclsnofoot]{IEEEtran}%

\usepackage{amsbsy}
\usepackage{floatflt} 

\usepackage{amsmath}
\usepackage{amssymb}
\usepackage{times}
\usepackage{graphicx}
\usepackage{xspace}
\usepackage{paralist} 
\usepackage{setspace} 
\usepackage{xypic}
\xyoption{curve}
\usepackage{latexsym}
\usepackage{theorem}
\usepackage{ifthen}
\usepackage{subfigure}

\usepackage{booktabs}
{\theoremheaderfont{\it} \theorembodyfont{\rmfamily}
\newtheorem{theorem}{Theorem}

\newtheorem{definition}[theorem]{Definition}

\newtheorem{corollary}[theorem]{Corollary}

\newtheorem{result}[theorem]{Result}

}

\def\ln{{\rm ln}}

\title{Topology for Distributed Inference on Graphs}
\author{Soummya Kar, Saeed Aldosari, and Jos\'e M.~F.~Moura$^{*}$
\thanks{The 1st and 3rd authors are with the Dep.~ECE,
Carnegie Mellon University, Pittsburgh, PA, USA 15213 (e-mail:
\{soummyak,moura\}@ece.cmu.edu, ph: (412)268-6341, fax: (412)268-3890.) The 2nd author is with EE Dept., King Saud University, P.~O.~Box 800, Riyadh, 11412, Saudi Arabia, (dosari@ksu.edu.sa, ph: +966-553367274, fax: + 966-1-4676757.)}
\thanks{Work supported by the DARPA DSO Advanced Computing and Mathematics Program Integrated Sensing and Processing (ISP) Initiative under  ARO grant \#~DAAD19-02-1-0180 and by NSF under grants \#~ECS-0225449 and~\#~CNS-0428404.}}

\begin{document}
\maketitle \thispagestyle{empty}
\maketitle
\begin{abstract}
Let~$N$ local decision makers in a sensor network  communicate with their neighbors to reach a decision \emph{consensus}.  Communication is local,  among neighboring sensors only, through noiseless or noisy links. We study the design of the  network topology that optimizes the rate of convergence of the iterative decision consensus algorithm. We reformulate the topology design problem as a spectral graph  design problem, namely, maximizing the eigenratio~$\gamma$ of two eigenvalues of the graph Laplacian~$L$, a matrix that is naturally associated with the interconnectivity pattern of the network. This reformulation avoids costly Monte Carlo simulations  and leads to the class of non-bipartite Ramanujan graphs for which we find a lower bound on~$\gamma$. For Ramanujan topologies and noiseless links, the local probability of error converges much faster to the overall global probability of error than  for structured graphs, random graphs, or graphs exhibiting small-world characteristics. With noisy links, we determine the optimal number of iterations before calling a decision. Finally,  we introduce a new class of random graphs  that are easy to construct, can be designed with arbitrary number of sensors,  and whose spectral and convergence properties make them practically equivalent to  Ramanujan topologies.
\end{abstract}

\noindent{\bf EDICS:} SEN-DIST, SEN-FUSE
\newpage
\section{Introduction}
\label{introduction}
The paper studies the problem of designing the topology of a graph network. As a motivational application we consider the problem of describing the connectivity graph of a sensor network, i.e., specifying  with which sensors should each sensor in the network communicate.  We will show that the topology of the network has a major impact on the convergence of distributed inference algorithms, namely, that these algorithms converge much faster for certain connectivity patterns than for others, thus requiring  much less intersensor communication and power expenditure. 

The literature on topology design for distributed detection is scarce. Usually, the underlying communication graph is specified ab initio as a  structured graph, e.g., parallel networks where sensors communicate with a fusion center, \cite{SensNets:Tenney81,SensNets:Tsitsiklis88,SensNets:Tsitsiklis93,SensNets:Willett00}, or serial networks where communication proceeds  sequentially from a sensor to the next; for these and other similar architectures, see Varshney~\cite{SensNets:Varshney} or \cite{SensNets:Blum97,SensNets:Chamberland03}.  These networks may not be practical; e.g., a parallel network depends on the integrity of the fusion center. 

We  published preliminary results on topology design for distributed inference problems in~\cite{SensNets:AldosariAsilomar05,aldosarimouramay06}. We  restricted the class of topologies to structured graphs,  random graphs obtained with the Erd\"os-R\'enyi construction, \cite{erdosrenyi59,erdosrenyi60,erdosrenyi61}, see also \cite{gilbert59,SensNets:Bollobas98}, or   random constructions  that exhibit small-world characteristics, see Watts-Strogatz~\cite{SensNets:Watts98}, see also Kleinberg~\cite{SensNets:KleinbergNature2000,SensNets:KleinbergThComp2000}. 
 We considered tradeoffs among these networks,  their number of links~$M$, and the number of bits~$b$ quantizing the state of the network at each sensor, under a global rate constraint, i.e.,  $Mb=K$, $K$ fixed. We adopted as criterion the convergence of the average probability of error $P_e$, which required extensive simulation studies to find the desired network topology. Reference~\cite{SensNets:Olfati04}  designs  Watts-Strogatz topologies in distributed consensus estimation problems, adopting as criterion the algebraic connectivity $\lambda_2(L)$ of the graph.


This paper designs good topologies for sensor networks, in particular,  with respect to the rate of convergence of   \emph{iterative} consensus  and \emph{distributed} detection algorithms. We consider the two cases of noiseless and noisy network links. We assume that   the total number~$M$ of communication links between sensors is fixed and that the graph weights are uniform across all network links. 
 This paper shows that, for both the iterative average-consensus and the distributed detection problems, the topology design  problem   is equivalent to the problem of maximizing with respect to the network topology  a certain graph spectral parameter~$\gamma$. This parameter is the ratio of the  algebraic connectivity  of the  graph over the largest eigenvalue~$\lambda_N(L)$ of the graph Laplacian~$L$. The algebraic connectivity of a graph, terminology introduced by \cite{SensNets:Fiedler73},  is the second smallest eigenvalue $\lambda_2(L)$ of its discrete Laplacian; see section~\ref{preliminaries}, the Appendix, and reference \cite{FanChung} for the definition of relevant spectral graph concepts. With this reinterpretation, we show that
the class of  Ramanujan graphs  essentially provides the optimal network topologies, exhibiting remarkable convergence properties, orders of magnitude faster than other structured or random small-world like networks. When the links are noisy, our analysis determines what is the optimal number of iterations to declare a decision. Finally,
 we present a new class  of random regular graphs whose performance is very close to the performance of Ramanujan graphs. These graphs can be designed with arbitrary number of nodes,  overcoming the limitation that the available constructions of Ramanujan graphs are restricted to networks whose number of sensors are limited to a sparse subset of the integers. 

We now summarize the paper. Section~\ref{preliminaries} and the Appendix 
  recall basic concepts and results from algebra and spectral graph theory. Section~\ref{sec:consensusalgorithm} presents the optimal \emph{equal} weights consensus algorithm and establishes its convergence rate  in terms of a spectral parameter. Section~\ref{ramanujan} defines formally the topology design problem, shows that Ramanujan graphs provide essentially the optimal topologies, and presents explicit algebraic constructions available in the literature for the Ramanujan graphs.  Section~\ref{distributedinference} considers distributed inference and shows that the average-consensus algorithm with noiseless links achieves asymptotically the optimal detection performance---that of a parallel architecture with a fusion center. This section shows that with noisy communication links there is an optimal maximum number of iterations to declare a decision. Section~\ref{results} demonstrates with several experiments the superiority   of the Ramanujan designs over other different alternative topologies, including structured networks, Erd\"os-Ren\'{y}i random graphs, and small-world type topologies. Section~\ref{randomgraphscloseramanujan} presents the new class of random regular Ramanujan like graphs that are easy to design with arbitrary number of sensors and that exhibit convergence properties close to Ramanujan topologies. Finally, section~\ref{conclusion} concludes the paper.

\section{Algebraic Preliminaries}
\label{preliminaries}


{\bf Graph Laplacian}
\label{graphlaplacian}
The topology of the sensor network is given by an undirected graph $G=(V,E)$, with nodes $v_{n}\in V$, $n\in \mathcal{I}=\{ 1,...,N\}$, and edges the unordered pairs  $e=
(v_{n},v_{l})\in E$, or, simply, $e$ = $(n,l)$, where $v_{n}$ and $v_{l}$
are called the edge endpoints. The edge $e=(n,l)\in E$ whenever sensor $v_{n}$ can communicate with $v_{l}$,
in which case the vertices $v_{n}$ and $v_{l}$ are
 adjacent and we write
$v_{n}\sim v_{l}$.  

\label{algebraicgraphtheory}
 We assume that the cardinality of $E$ is $|E|=M$ and, when needed, label the edges by $m$, $m=1,\cdots,M$. The terms sensor, node, and vertex are assumed to be
equivalent in this paper. 
A loop is an edge whose endpoints are
the same vertex. Multiple edges are edges with the same pair of endpoints. A
graph is \emph{simple} if it has no loops or multiple edges. A graph with loops or multiple edges is called a multigraph. A \emph{path} is a sequence $v_{n_0},\cdots, v_{n_{m}}$  such that $e_l=(v_{n_{l-1}},v_{n_{l}})\in E$, $l=1,\cdots,m$, and the $v_{n_l}$, $l=0,\cdots,m-1$, are all distinct.
 A graph is \emph{connected} if there is a
path from every sensor $v_{n}$ to every other sensor
$v_{l}$, $n,l=1,\ldots,N$. In this paper we assume the graphs to be \emph{simple} and \emph{connected},
unless otherwise stated. 

We can assign to a graph an $N\times N$
adjacency matrix $A$ (where, we recall, $N=|V|$,) defined by
\begin{equation}
\label{adjacency}
a_{n,l} = \left\{ \begin{array}{ll}
                    1 & \mbox{if $(n,l) \in E$} \\
                    0 & \mbox{otherwise}
                   \end{array}
          \right.
\end{equation}
The set of neighbors of  node~$n$ is 
\mbox{$\Omega_{n} = \{l:(n,l) \in E\}$} and its degree, $\mbox{deg}(n)$, is
the number of its neighbors, i.e., the cardinality $|\Omega_{n}|$. 
A graph is 
$k$-\emph{regular} if all vertices have the same degree~$k$.

The degree
matrix, $D$ is the $N\times N$ diagonal matrix $D=\mbox{diag}\left[d_{1,1}\cdots d_{N,N}\right]$  defined by
\begin{equation}
\label{degreeD}
d_{n,n}=\mbox{deg}(n)
\end{equation}
The Laplacian~$L$ of the graph, \cite{FanChung}, is the $N \times
N$ matrix defined by
\begin{equation}
\label{laplacian}
L = D - A
\end{equation}

{\bf Spectral properties of graphs.}
\label{spectrum}
We consider spectral properties of  \emph{connected regular} graphs.  Since the adjacency matrix $A$ is symmetric, all its eigenvalues
are real. Arrange the eigenvalues
of the adjacency matrix~$A$ as,
\begin{equation}
\label{eigenvaluesA}
k=\lambda_{1}(A)> \lambda_{2}(A)\geq \ldots \geq \lambda_{N}(A)\geq
-k
\end{equation}
It can be shown that the multiplicity of the largest eigenvalue~$\lambda_1(A)=k$
equals the number of connected components in the graph.
Then, for a connected graph, the multiplicity of the
eigenvalue $\lambda_{1}(A)=k$ is 1, which explains the strict inequality on the left in~(\ref{eigenvaluesA}). Also, $-k$ is an eigenvalue of~$A$ \emph{iff} the
graph is bipartite (please refer to the Appendix for the
definition of bipartite graphs.) Hence, for non-bipartite graphs,
$\lambda_{N}(A)>-k$. In this paper, we focus on connected, non-bipartite graphs.

The Laplacian is a symmetric, positive
semi-definite matrix, and, consequently, all its eigenvalues are
non-negative.   It follows from~(\ref{eigenvaluesA}) that the eigenvalues of the Laplacian satisfy
\begin{equation}
\label{eigenvaluesL}
0=\lambda _{1}(L)<\lambda _{2}(L) \leq ...\leq \lambda _{N}(L)
\end{equation}
The multiplicity of the zero eigenvalue of~$L$ equals the number of
connected components in the graph, which explains the strict
inequality on the left hand side of~(\ref{eigenvaluesL}) for the case of connected graphs. For
$k$-regular graphs, the eigenvalues of~$A$ and~$L$ are directly related by
\begin{equation}
\label{eig_rel}
\forall n\in \mathcal{I}:\:\:\: \lambda_{n}(L)=k-\lambda_{n}(A)
\end{equation}

We write the eigendecomposition of the Laplacian~$L$ as
\begin{eqnarray}
\label{L-eigendecomposition-1}
L&=&\mathbf{U}\Lambda\mathbf{U}^T\\
\label{L-eigendecomposition-2}
&=&\left[\mathbf{u}_1\cdots\mathbf{u}_N\right]
	\mbox{diag}\left[\lambda_1(L)\cdots\lambda_N(L)\right]
		\left[\mathbf{u}_1\cdots\mathbf{u}_N\right]^T
\end{eqnarray}
where the $\mathbf{u}_n$, $n=1,\cdots,N$,  are orthonormal and $\mbox{diag}\left[\cdots\right]$ is a diagonal matrix. 
 We note that, from the structure of~$L$, each diagonal entry of~$D$ is the corresponding row sum of~$A$, so the eigenvector $\mathbf{u}_1$ corresponding to the zero eigenvalue $\lambda_1(L)$ is the (normalized) vector of one's 
\begin{equation}
\label{eq:u1}
\mathbf{u}_1=\frac{1}{\sqrt{N}}\mathbf{1}=\frac{1}{\sqrt{N}}\left[1\cdots1\right]^T
\end{equation}

\section{Average Consensus Algorithm}
\label{sec:consensusalgorithm}
\label{distributeddetection}
\label{problemstatement}
In this Section, we present the consensus algorithm in Subsection~\ref{sub:consensusalgorithm}, discuss the case of equal weights in Subsection~\ref{linkweights}, and  establish the convergence rate of the algorithm in Subsection~\ref{sub:convergencerate}.
\label{consensusalgorithm}
\subsection{Consensus Algorithm Description} 
\label{sub:consensusalgorithm}
We review briefly the consensus algorithm that computes in a distributed fashion the average of $N$ quantities $r_n$, $n=1,\cdots,N$.  Assume a sensor network with interconnectivity graph $G=(V,E)$ defined by a neighborhood system $\Omega=\{\Omega_n: n\in \mathcal{I}\}$, and where $\Omega_n$ is the set of neighbors of sensor~$n$.
Initially, sensors take measurements $r_{1},\ldots r_{N}$. It is desired to compute their mean in a distributed fashion, 
\begin{equation}
\label{meanr}
\overline{r}=\frac{1}{N}\sum_{n=1}^N r_n
\end{equation}
i.e., by  only local communication among  neighbors.
Define 
the \emph{state} at iteration $i=0$ at sensor~$n$ by
\[
x_{n}(i=0)=r_{n},\:\:n=1,\cdots,N
\]
Iterative consensus is  carried out according to the following linear
operation, \cite{SensNets:Xiao04},
\begin{equation}
x_{n}(i)=W_{nn}x_{n}(i-1)+\underset{l\in\Omega_{n}}{\overset{N}{\sum}}%
W_{nl}x_{l}(i-1) \label{Eqn: Update basic}
\end{equation}
where 
$W_{nl}$ is a weight associated with edge $(n,l)$, if this edge
exists. 
The weight $W_{nl}=0$, $n\neq l$,  when there is no link associated with
it, i.e.,  if
$(n,l)\notin E
$. 
 The value $x_n(i)$ stored at iteration~$i$ by sensor~$n$ is the \emph{state} of $v_n$ at~$i$.  The consensus~(\ref{Eqn: Update basic}) can be expressed in matrix form as
\begin{equation}
\mathbf{x}_i=W\mathbf{x}_{i-1}
\label{Eqn: Update t}%
\end{equation}
where $\mathbf{x}_i$ is the $N\times1$ vector of all current
states and $W=\left[W_{nl}\right]$ is the matrix of all the weights.
 The updating~(\ref{Eqn: Update t}) can be written in terms
of the initial states as%
\begin{eqnarray}
\mathbf{x}_i&=&W^{i}\mathbf{x}_0
\label{Eqn:Update x0}\\
\label{xinitial}
\mathbf{x}_0&=&\left[x_1(0)\cdots x_N(0)\right]^T=\left[r_1\cdots r_N\right]^T
\end{eqnarray}

Let the $N$-dimensional vector $\mathbf{1}=[1\cdots 1]^T$. Convergence to consensus occurs if
\begin{eqnarray}
\nonumber
\forall n:\:\:\lim_{i\rightarrow\infty}x_{n}(i)&=& \overline{r}\\
\label{def:x_global}
\lim_{i\rightarrow\infty}\mathbf{x}_i&=&\overline{\mathbf{x}}=\overline{r}\,\,\mathbf{1}\\
\label{consensusconvergenceW}
\lim_{i\rightarrow\infty}W^i&=&\frac{\mathbf{1}\mathbf{1}^T}{N}
\end{eqnarray}
\subsection{Link Weights}
\label{linkweights} \label{Sec: WeightDesign} 
  The convergence
speed of the  iterative consensus  depends on the choice of the link
weights, $W_{nl}$. 
 In this paper, we consider only the case of 
\emph{equal} weights, i.e., we assign an equal weight $\alpha$ to
all network links.  
 $I$ and $L$ be the $N$-dimensional identity matrix and the graph Laplacian. The weight matrix becomes
\begin{equation}
\label{iter}
W=I-\alpha L
\end{equation}
For a particular network topology, the value of~$\alpha$ that maximizes the
convergence speed is, \cite{SensNets:Xiao04}, %
\begin{equation}
\alpha^{\ast}=\frac{2}{\lambda_{2}(L)+\lambda_{N}(L)}%
\label{Eqn: opt const weights}%
\end{equation}
 For proofs of these statements and other
weight design techniques, the reader is referred to \cite{SensNets:Xiao04} and \cite{SensNets:Olfati04}.

We now consider the eigendecomposition of the weight matrix~$W$. From~(\ref{iter}), with the optimal weight~(\ref{Eqn: opt const weights}), using the eigendecomposition~(\ref{L-eigendecomposition-2}) of~$L$, we have that
\begin{eqnarray}
\label{eq:Wexpansion-1}
W&=&\left[\mathbf{u}_1\cdots\mathbf{u}_N\right]\mbox{diag}\left[\gamma_1\cdots\gamma_N\right]\left[\mathbf{u}_1\cdots\mathbf{u}_N\right]^T\\
\label{eq:Wexpansion}
&=&\sum_{n=1}^N\gamma_n\mathbf{u}_n\mathbf{u}_n^T,
\end{eqnarray}
where: 
$\mathbf{u}_n$, $n=1,\cdots,N$,  are  the  orthonormal eigenvectors of~$L$, and \`a fortiori of~$W$; and
$\mbox{diag}\left[\gamma_1\cdots\gamma_N\right]$ is the diagonal matrix of the eigenvalues $\gamma_n$ of~$W$. These eigenvalues are 
\begin{equation}
\label{gamman}
\gamma_n=1-\alpha^*\lambda_n(L)
\end{equation}
From  the spectral properties of the Laplacian of a connected graph, and the choice of $\alpha^*$, the eigenvalues of~$W$ satisfy
\begin{eqnarray}
\label{gamma2gamman-2}
&1=\gamma_1>\gamma_2\geq\cdots\geq\gamma_N&\\
\label{gamma2gamman}
\forall n>1:&\left|\gamma_n\right|\leq\gamma_2<1&
\end{eqnarray}


\subsection{Consensus Algorithm: Convergence Rate} 
\label{sub:convergencerate}
We now study the convergence rate of the consensus algorithm.

\begin{result}
\label{consensusalg:convergencerate}
For any connected graph~$G$, the convergence rate of the consensus algorithm~(\ref{Eqn: Update t}) or~(\ref{Eqn:Update x0}) is
\begin{equation}
\label{norm_bound-1}
\|\mathbf{x}_{i}-\overline{\mathbf{x}}\|\leq\|\mathbf{x}_{0}-\overline{\mathbf{x}}\| \gamma_{2}^i
\end{equation}
where $\overline{\mathbf{x}}$ and $\mathbf{x}_{0}$ are given in~(\ref{def:x_global}) and~(\ref{Eqn:Update x0}) and
\begin{eqnarray}
\label{norm-bound-2}
\gamma_2&=&\frac{1-\gamma}{1+\gamma}\\
\label{norm-bound-3}
\gamma&=&\frac{\lambda_2(L)}{\lambda_N(L)}
\end{eqnarray}
\end{result}
\begin{proof}
Represent the vector $\mathbf{x}_0$ in~(\ref{xinitial}) in terms of the eigenvectors $\mathbf{u}_n$ of~$L$
\begin{equation}
\label{eig_dec}
\mathbf{x}_0=\sum_{n=1}^{N} d_{n} \mathbf{u}_{n}
\end{equation}
where $d_{n}=\mathbf{x}_0^T\mathbf{u}_{n}$. From the value of~$\mathbf{u}_1$ in~(\ref{eq:u1})
 it follows that
\begin{equation}
\label{eq:d1}
d_{1}=
\sqrt{N}\overline{r}
\end{equation}
 Replacing~(\ref{eq:Wexpansion})
and~(\ref{eig_dec}) in~(\ref{Eqn:Update x0}) and using~(\ref{eq:d1}) and the orthonormality of the eigenvectors of~$L$ (and~$W$,) we obtain
\begin{eqnarray}
\mathbf{x}_{i} & = & W^{i}\,\mathbf{x}_{0} \nonumber \\
		&=&\sum_{l=1}^N\gamma^i_l\mathbf{u}_l\mathbf{u}_l^T \sum_{n=1}^{N} d_{n} \mathbf{u}_{n}\\
\nonumber
		&=&\sum_{n,l=1}^N d_n\gamma^i_l\mathbf{u}_l\mathbf{u}_l^T \mathbf{u}_{n}\\
\nonumber
               & = & \sum_{l=1}^{N}d_{l}\,\gamma_{l}^{i}\,\mathbf{u}_{l} \\
\nonumber
		&=&d_1\gamma^i_1\mathbf{u}_1+ \sum_{l=2}^{N}d_{l}\,\gamma_{l}^{i}\,\mathbf{u}_{l}\\
\label{eq:xixbar}
               & = & \overline{\mathbf{x}}   
+\sum_{l=2}^{N}d_{l}\,\gamma_{l}^{i}\,\mathbf{u}_{l}
\end{eqnarray}
where $\overline{\mathbf{x}}$ is given in
eqn~(\ref{def:x_global}). From these it follows that
\begin{eqnarray}
\|\mathbf{x}_{i}-\overline{\mathbf{x}}\| & = &
\left\|\sum_{l=2}^{N}d_{l}\gamma_{l}^{i}\mathbf{u}_{l}\right\| \nonumber \\
                                     & \leq &
                                     \left|\gamma_{2}^{i}\right|\,\,\left\|\sum_{l=2}^{N}d_{l}\mathbf{u}_{l}\right\|
\label{bound1-a}                                      \\
\label{bound1-b}
                                     & = &
                                     \left\|\mathbf{x}_{0}-\overline{\mathbf{x}}\right\|\,\,
				\gamma_{2}^{i}
\end{eqnarray}
To get~(\ref{bound1-a}), we used the bounds given by~(\ref{gamma2gamman}). To obtain~(\ref{bound1-b}) we used the fact that from~(\ref{eq:xixbar}), for $i=0$, it follows that  
\begin{equation}
\label{norm_bound}
\left\|\sum_{l=2}^{N}d_{l}\gamma_{l}^{i}\mathbf{u}_{l}\right\| =\left\|\mathbf{x}_{0}-\overline{\mathbf{x}}\right\|
\end{equation}
 From~(\ref{norm_bound}) it follows that, to obtain the optimal convergence
rate, $\gamma_{2}$ should be as small as possible. From the expression for $\gamma_n$ in~(\ref{gamman}), and using the optimal choice for~$\alpha$ in~(\ref{Eqn: opt const weights}), we get successively
\begin{eqnarray}
\gamma_{2} & = & 1-\alpha \lambda_{2}(L) \\ \nonumber
           & = & 1-\frac{2\lambda_{2}(L)}{\lambda_{2}(L)+\lambda_{N}(L)} \\ \nonumber
           & = & \frac{\lambda_{N}(L)-\lambda_{2}(L)}{\lambda_{N}(L)+\lambda_{2}(L)} \\
\label{gamma2-2}
           & = & \frac{1-\lambda_{2}(L)/\lambda_{N}(L)}{1+\lambda_{2}(L)/\lambda_{N}(L)}
\end{eqnarray}
Thus, the minimum value of $\gamma_{2}$ is attained when the ratio
\begin{equation}
\label{gammaratio}
\lambda_{2}(L)/\lambda_{N}(L)
\end{equation}
 is maximum, i.e.,
\begin{equation}
\label{gammalambdaoptimization}
\max \mbox{convergence rate}\sim\min\gamma_2\sim\max\gamma=\max\frac{\lambda_2(L)}{\lambda_N(L)}
\end{equation}
\end{proof}




\section{Topology Design: Ramanujan Graphs}
\label{performancedifferenttopologies}
\label{ramanujan}
In this section, we consider the problem of designing the topology of a sensor network that maximizes the rate of convergence of the average consensus algorithm. Using the results of Section~\ref{consensusalgorithm},  in Subsection~\ref{Prob-Red}, we reformulate the average consensus topology design as a spectral graph topology design problem by  restating it in terms of the design of  the topology of the network that maximizes  an eigenratio of two eigenvalues of the graph Laplacian, namely, the graph parameter~$\gamma$ given by~(\ref{norm-bound-3}). We then consider in Subsection~\ref{ramanujangraphs} the class of Ramanujan graphs and show in what sense they are good topologies. Finally, Subsection~\ref{ramanujanconstruction} describes algebraic constructions of Ramanujan graphs available in the literature, see~\cite{LPS}.
\subsection{Topology Optimization} 
\label{Prob-Red}
We formulate the design of the topology of the sensor network for the average consensus algorithm as the optimization of the spectral eigenratio parameter~$\gamma$, see~(\ref{norm-bound-3}).
From our  discussion in Section~\ref{consensusalgorithm}, it follows that
the topology that optimizes the convergence rate of the consensus algorithm can be restated as the following graph optimization problem:
\begin{equation}
\label{spectralgraphtopologydesign}
\max_{G\in \mbox{ } \mathcal{G}}\gamma=\max_{G\in \mbox{ } \mathcal{G}} \frac{\lambda_{2}(L)}{\lambda_{N}(L)}
\end{equation}
where $\mathcal{G}$ denotes the set of all possible simple connected graphs
with~$N$ vertices and~$M$ edges. 

We remark that~(\ref{spectralgraphtopologydesign}) will be significant because we will be able to use spectral properties of graphs to propose a class of graphs---the Ramanujan graphs---for which we can present a lower bound on the spectral parameter~$\gamma$. This avoids the lengthy and costly Monte Carlo simulations used to evaluate the performance of other topologies as done, for example, in our previous work, see~\cite{SensNets:AldosariAsilomar05,aldosarimouramay06} or in~\cite{SensNets:Olfati04}.

\subsection{Ramanujan Graphs}
\label{ramanujangraphs}
In this section, we consider $k$-regular graphs.
Before introducing the class of Ramanujan graphs, we discuss several bounds on eigenvalues of graphs. We first  state a
well-known result from algebraic graph theory.
\begin{theorem}[Alon and Boppana~\cite{Alon,LPS}]
\label{Result_Alon}
 Let $G = G_{N,k}$ be a $k$-regular graph on $N$
vertices. Denote by $\lambda_{G}(A)$, the absolute value of the
largest eigenvalue (in absolute value) of the adjacency matrix
$A$ of the graph~$G$, which is distinct from $\pm k$; in other words,
$\lambda^{2}_{G}(A)$ is the next to largest eigenvalue of $A^{2}$.
Then
\begin{equation}
\label{eqn:resultalon}
\liminf_{N\rightarrow \infty}\lambda_{G}(A) \geq 2\sqrt{k-1}
\end{equation}
\end{theorem}

A second result, \cite{Alon}, also shows 
that, for an infinite
family of $k$-regular graphs $G_{m}$, $m\in \{1,2,\cdots\}$, for which
the number of nodes diverges as~$m$ becomes large, the algebraic connectivity~$\lambda_{2}(L)$
of the graphs is asymptotically bounded by
\begin{equation}
\label{lambda_2_bound}
 \liminf_{N\rightarrow \infty}\lambda_{2}(L) \leq k-2\sqrt{k-1}
\end{equation}
Note that~(\ref{lambda_2_bound}) is a direct upperbound on the limiting behavior of $\lambda_2(L)$ itself, while from~(\ref{eqn:resultalon}) we may derive an upperbound on the limiting behavior of $\lambda_2(A)$ or of $\lambda_N(A)$, depending if $\lambda_2(A)\leq\left|\lambda_N(A)\right|$ or $\lambda_2(A)\geq\left|\lambda_N(A)\right|$ in the limit. We consider each of these two cases separately.
%
%
%

\begin{enumerate}
\item $\liminf_{N\rightarrow \infty}\lambda_2(A)\leq\liminf_{N\rightarrow \infty}\left|\lambda_N(A)\right|:
\liminf_{N\rightarrow \infty}|\lambda_{N}(A)|\geq
2\sqrt{k-1}.$

Since $\lambda_{N}(A)\leq0$, it follows that for $k$-regular connected simple graphs
\[
\liminf_{N\rightarrow \infty}\lambda_{N}(A)\leq -2\sqrt{k-1}
\]
From this,  we have
\begin{equation}
\label{caseIlambdaN} 
\liminf_{N\rightarrow \infty}\lambda_{N}(L)\geq
k+2\sqrt{k-1}
\end{equation}
Combining~(\ref{caseIlambdaN}) with~(\ref{lambda_2_bound}), we get using standard results from limits of series of real numbers
\begin{equation}
\label{casIlambdaratiobound} \liminf_{N\rightarrow
\infty}\gamma(N)=\liminf_{N\rightarrow
\infty}\frac{\lambda_{2}(L)}{\lambda_{N}(L)}\leq
\frac{k-2\sqrt{k-1}}{k+2\sqrt(k-1)}
\end{equation}
Eqn~(\ref{casIlambdaratiobound}) is an asymptotic upper bound on the spectral eigenratio parameter
$\gamma=\lambda_{2}(L)/\lambda_{N}(L)$ for the family of
non-bipartite graphs for which $\liminf\lambda_2(A)\leq\liminf\left|\lambda_N(A)\right|$.
\item  $\liminf_{N\rightarrow \infty}\lambda_2(A)\geq\liminf_{N\rightarrow \infty}\left|\lambda_N(A)\right|: \liminf_{N\rightarrow \infty}|\lambda_{N}(A)|\leq 2\sqrt{k-1}.$

Now Theorem~(\ref{Result_Alon}) is inconclusive with respect to $\liminf_{N\rightarrow
\infty}\lambda_N(A)$. From the fact that $-k\leq\lambda_N(A)\leq 0$, we can promptly deduce that
$k\leq\lambda_N(L)\leq 2k$. Combining this with~(\ref{lambda_2_bound}), we get
\begin{equation}
\label{casIIlambdaratiobound} \liminf_{N\rightarrow
\infty}\frac{\lambda_{2}(L)}{\lambda_{N}(L)}\leq \frac{k-2\sqrt{k-1}}{k}
\end{equation}
which gives an asymptotic upper bound for the eigenratio parameter
$\gamma=\lambda_{2}(L)/\lambda_{N}(L)$ for the family of
non-bipartite  graphs satisfying  $\liminf\lambda_2(A)\geq\liminf\left|\lambda_N(A)\right|$.
\end{enumerate}

%
%
%

We now consider the class of Ramanujan graphs.

\begin{definition}[Ramanujan Graphs]
\label{def:Ram}
 A graph $G = G_{N,k}$ will be called Ramanujan if
\begin{equation}
\label{ramanujan1}
\lambda_{G}(A) \leq 2\sqrt{k-1}
\end{equation}
\end{definition}
Graphs with small $\lambda_{G}(A)$ (often called graphs with large
spectral gap in the literature) are called expander graphs, and
the Ramanujan graphs are one of the best explicit expanders
known. 
Note that Theorem~\ref{Result_Alon} and~(\ref{eqn:resultalon}) show that, for general graphs, $\lambda_G(A)$ is in the limit  \emph{lower} bounded  by $2\sqrt{k-1}$, while for Ramanujan graphs $\lambda_G(A)$ is, for every finite $N$, \emph{upper} bounded by $2\sqrt{k-1}$. 

From~(\ref{ramanujan1}), it follows that, for non-bipartite Ramanujan graphs, 
\begin{eqnarray}
\label{ramanujan2}
\lambda_{2}(A)
&\leq& \phantom{-}2\sqrt{k-1}\\
\label{ramanujan3}
\lambda_{N}(A)&\geq& -2\sqrt{k-1}
\end{eqnarray}
Equations~(\ref{ramanujan2}) and~(\ref{ramanujan3})
together with eqn~(\ref{eig_rel}) give, for non-bipartite
Ramanujan graphs,
\begin{eqnarray*}
\lambda_{2}(L)&\geq& k-2\sqrt{k-1}\\
\lambda_{N}(L)&\leq& k+2\sqrt{k-1}
\end{eqnarray*}
and, hence, for non-bipartite Ramanujan graphs
\begin{equation}
\label{Ram_bound}
\gamma=\frac{\lambda_{2}(L)}{\lambda_{N}(L)}\geq \frac{k-2\sqrt{k-1}}{k+2\sqrt{k-1}}
\end{equation}
This is a key result and shows that for non-bipartite Ramanujan graphs the eigenratio parameter~$\gamma$ is lower bounded by~(\ref{Ram_bound}). It will explain in what sense we take Ramanujan graphs to be ``optimal'' with respect to the topology design problem stated in Subsection~\ref{Prob-Red} as we discussed next. To do this, we compare the lower bound~(\ref{Ram_bound}) on~$\gamma$ for Ramanujan graphs with the asymptotic upper bounds~(\ref{casIlambdaratiobound}) and~(\ref{casIIlambdaratiobound}) on~$\gamma$ for generic graphs. We consider the two cases separately again.
\begin{enumerate}
\item Generic graphs for which $\liminf_{N\rightarrow \infty}\lambda_2(A)\leq\liminf_{N\rightarrow \infty}\left|\lambda_N(A)\right|.$ Here,  the lower bound on~(\ref{Ram_bound})  and the upper bound on~(\ref{casIlambdaratiobound}) are the same. Since for any value of~$N$, (\ref{Ram_bound}) shows that $\gamma$ is above the bound, we conclude that, in the limit of  large~$N$, the eigenratio parameter~$\gamma$ for non-bipartite Ramanujan graphs approaches the bound from above. This contrasts with non-bipartite non-Ramanujan graphs for which in the limit of large~$N$ the eigenratio parameter~$\gamma$ stays below the bound. 
\item Generic graphs for which $\liminf_{N\rightarrow \infty}\lambda_2(A)\geq\liminf_{N\rightarrow \infty}\left|\lambda_N(A)\right|.$ Now the bound~(\ref{casIIlambdaratiobound}) does not help in asserting that Ramanujan graphs have faster convergence than these generic graphs. This is because
\[
\frac{k-2\sqrt{k-1}}{k+2\sqrt{k-1}}<\frac{k-2\sqrt{k-1}}{k}
\] 
i.e., the lower bound~(\ref{Ram_bound}) for Ramanujan graphs is smaller than the upper bound~(\ref{casIIlambdaratiobound}) for generic graphs. We should note that the ratio of two quantities is usually much more sensitive to variations in the numerator than to variations of the denominator. Because Ramanujan graphs optimize the algebraic connectivity of the graph, i.e., $\lambda_2(L)$, we still expect $\gamma$ to be much larger for Ramanujan graphs than for these graphs. We show in Section~\ref{results} this to be true for broad classes of graphs, including,  structured graphs, small-world graphs, and Erd\"{o}s-Ren\'{y}i random graphs.
\end{enumerate}
%
\subsection{Ramanujan graphs: Explicit Algebraic Construction}
\label{ramanujangraphsconstructions}
We now provide explicit constructions of Ramanujan graphs available in the literature.
We refer the reader  to the Appendix for the definitions
of the various terms used in this section. The explicit
constructions presented next are based
on the construction of Cayley graphs. The following paragraph
gives a brief overview of the Cayley graph construction.

{\bf Cayley Graphs.}
The Cayley graph construction gives a simple procedure for constructing
$k$-regular graphs using group theory. Let $X$ be a finite group
with $|X|=N$, and $S$ a $k$-element subset of $X$. For the graphs
used in this paper, we assume that $S$ is a symmetric subset of $X$,
in the sense that $s\in S$ implies $s^{-1}\in S$. We now construct a
graph $G = G(X,S)$ by having the vertex set to be the elements of
$X$, with $(u,v)$ as an edge if and only if $vu^{-1}\in S$. It can
be easily verified that, for a symmetric subset $S$, the graph
constructed above is $k$-regular on $|X|$ vertices. The subset $S$
is often called the set of generators of the Cayley graph $G$, over
the group $X$.
\label{ramanujanconstruction}
Explicit constructions of Ramanujan graphs for a fixed~$k$ and
varying $N$, \cite{Murty}, have been described for the cases
$k-1$ is a prime, \cite{LPS}, \cite{Margulis}, or a prime power,
~\cite{Morgenstern}. The Ramanujan graphs used in this paper are
obtained using the Lubotzky-Phillips-Sarnak (LPS) construction,
~\cite{LPS}. We describe two constructions of non-bipartite
Ramanujan graphs in this section, \cite{LPS}, and refer to them as
LPS-I and LPS-II, respectively.

{\bf LPS-I Construction.} We
consider two unequal primes $p$ and $q$, congruent to 1 modulo 4,
and further let the Legendre symbol $\left(\frac{p}{q}\right)=1$.
The LPS-I graphs are Cayley graphs over the PSL(2,$Z/qZ$) group
(Projective Special Linear group over the field of integers modulo
$q$.) (Precise definitions and explanations of these terms are
provided in the Appendix.) Hence, in this case, the group $X$ is
the PSL(2,$Z/qZ$) group. It can be shown that the number of
elements in $X$ is given by
\[
|X|=\frac{q(q^{2}-1)}{2},
\]
see~\cite{LPS}. To get the symmetric subset $S$ of generators,
we consider the equation,
\[
a_{0}^{2}+a_{1}^{2}+a_{2}^{2}+a_{3}^{2}=p,
\]
where $a_{0},a_{1},a_{2},a_{3}$ are integers. Let
\[
\beta = (a_{0},a_{1},a_{2},a_{3}),
\]
be a solution of the above equation. From a formula by Jacobi,
~\cite{jacobi_proof}, there are a total of $8(p+1)$ solutions of
this equation, and, out of them, $p+1$ solutions are such that
$a_{0}>0$ and odd, and $a_{j}$ even for $j=1,2,3$. Also, let $i$
be an integer satisfying
\[
i^{2}\equiv -1\bmod(q).
\]
For each of these $p+1$ solutions, $\beta$, we define the matrix
$\widetilde{\beta}$ in PSL(2,$Z/qZ$) as,
\begin{equation}
\widetilde{\beta} = \left( \begin{array}{rr}
                        a_{0}+ia_{1} & a_{2}+ia_{3} \\
                        -a_{2}+ia_{3} & a_{0}-ia_{1}
                        \end{array}
                \right)
\end{equation}
The Appendix shows that these $p+1$ matrices belong to the
PSL(2,$Z/qZ$) group. These $p+1$ matrices constitute the subset
$S$, and $S$ acts on the PSL(2,$Z/qZ$) group to produce the
$p+1$-regular Ramanujan graphs on $\frac{1}{2}q(q^{2}-1)$
vertices. The Ramanujan graphs thus obtained are non-bipartite,
see~\cite{LPS}. As an example of a LPS-I graph, we may choose
$p=17$ and $q=13$. We note that $p$ and $q$ are congruent to 1
modulo 4, and the Legendre symbol $\left(\frac{17}{13}\right)=1$.
The LPS-I graph with these values of $p$ and $q$ will be a regular
graph with
degree $k=p+1=18$ and has $\frac{q(q^{2}-1)}{2}=1092$ vertices.

The only problem with the LPS-I graphs is that the number of
vertices grows as $O(q^{3})$, which limits the use of such graphs.
In the next section the explicit construction of a second-class of
Ramanujan graphs is presented that avoids this difficulty. 

\begin{figure}
\begin{center}
\includegraphics[height=2in, width=2in ] {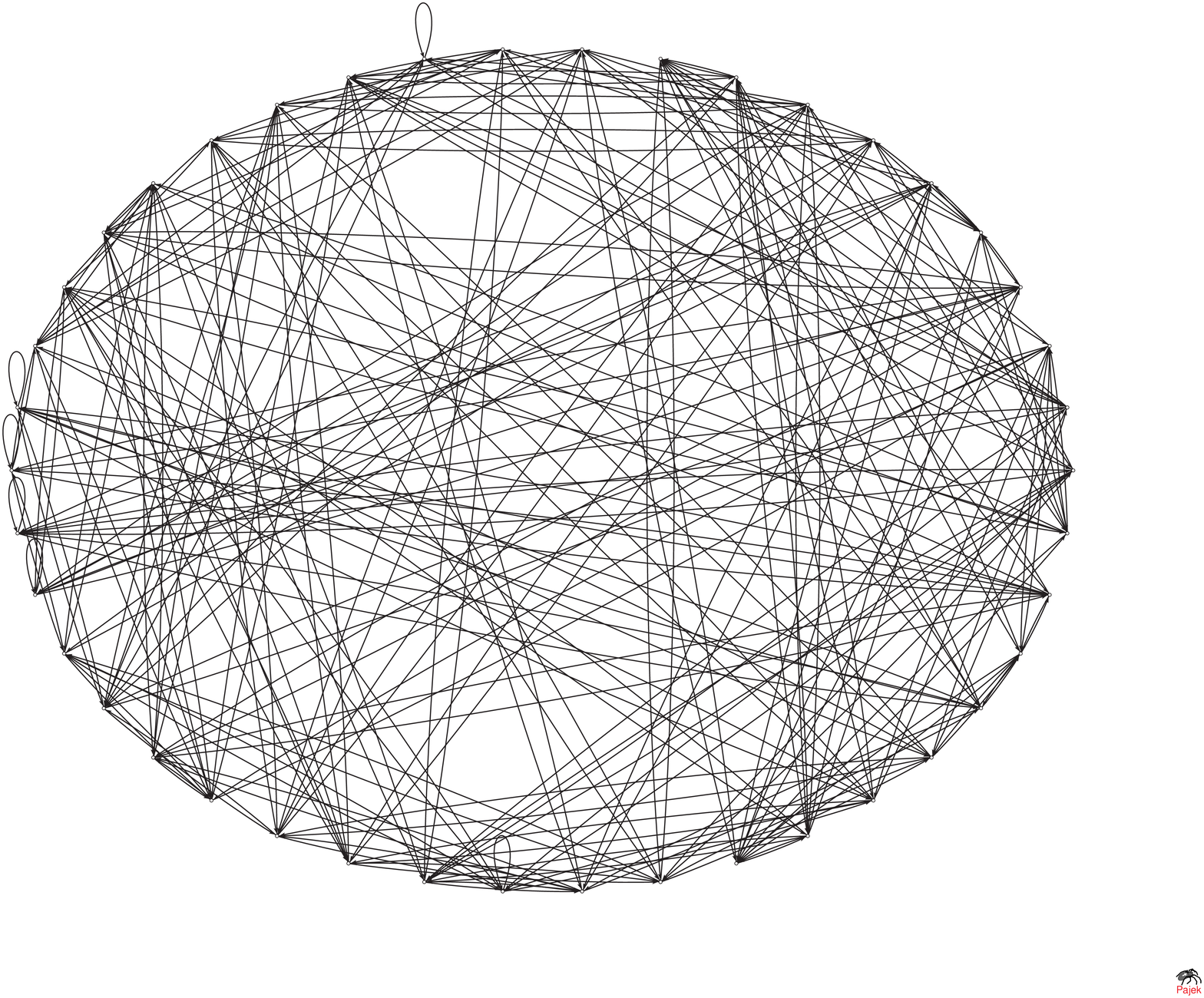} \caption{LPS-II
graph with number of vertices $N=42$ and degree $k=6$.}
\label{LPS-II_42}
\end{center}
\end{figure}

{\bf LPS-II Construction.} The LPS-II graphs are obtained in a slightly different
way. Here also, we start with two unequal primes $p$ and $q$
congruent to $1\bmod{4}$, such that the Legendre symbol
$\left(\frac{p}{q}\right)=1$. We define the set $P^{1}(F_{q}) =
\{0,1,...,q-1,\infty\}$, called Projective line over~$F_{q}$, and which is basically the set of integers
modulo $q$, with an additional ``infinite" element inserted in it.
It follows that $|P^{1}(F_{q})|=q+1$. The LPS-II graphs are
produced by the action of the set $S$ of the $p+1$ generators
defined above (LPS-I) on $P^{1}(F_{q})$, in a linear fractional
way. More information about linear fractional transformations is
provided in the Appendix. The Ramanujan graphs obtained in this
way, are non-bipartite $p+1$-regular graphs on $q+1$ vertices
~\cite{LPS}. The LPS-II graphs thus obtained, may few loops
~\cite{Lafferty}, which does not pose any problem because their
removal does not affect the Laplacian matrix and hence its
spectrum in any way (this is because the Laplacian $L=D-A$, and a
loop at vertex $n$ adds the same term to both $D_{nn}$ and
$A_{nn}$, which gets canceled while taking the difference.) The
LPS-II offers a larger family of Ramanujan graphs than LPS-I,
because in the former, the number of vertices grows only linearly
with $q$. As an example of a LPS-II Ramanujan graph, we take $p =
5$ and $q=41$. (It can be verified that $p,q\equiv 1\bmod(4)$ and
the Legendre symbol, $\left(\frac{p}{q}\right)=1$.) Thus, we have
a non-bipartite Ramanujan graph, which is 6-regular and has 42
vertices. Fig.~\ref{LPS-II_42} shows the graph, thus obtained.

\section{Distributed Inference}
\label{distributedinference}
In this Section, we apply the average-consensus algorithm to inference in sensor networks, in particular, to detection. This continues our work in~\cite{SensNets:AldosariAsilomar05,aldosarimouramay06}  where we compared  small-world topologies to Erd\"os-Ren\'{y}i random graphs and structured graphs. Subsection~\ref{sub:distributeddetection} formalizes the problem and Subsection~\ref{noiseanalysis} presents the noise analysis.
\subsection{Distributed Detection}
\label{sub:distributeddetection}
We study in this Section the simple binary   hypothesis test  where the state of the environment takes one of two possible alternatives, $H_{0}$ (target absent) or $H_{1}$ (target present). The true state~$H$ is monitored by a network~$G$ of~$N$ sensors. These collect
measurements $\mathbf{y}=(y_{1},\ldots y_{N})$ that are independent and identically
distributed (i.i.d.) conditioned on the  true state $H$; their known conditional probability density is $f_{i}(y)=f(y|H_{i})$, $i=0,1$. We first consider a parallel architecture where the sensors communicate to a single fusion center their local decisions.

Each sensor $v_{n}$, $n=1,\ldots,N$, starts by computing the (local) log-likelihood ratio (LLR)
\begin{equation}
\label{localdecision}
r_{n}=\ln\frac{\Pr(y_{n}|H_{1})}{\Pr(y_{n}|H_{0})}
\end{equation}
 of its measurement
$y_{n}$. 
 The local decisions are then transmitted to a fusion center. The central  decision is
\begin{equation}
\label{centraldecision}
\ell=\frac{1}{N}\sum_{n=1}^N r_n\overset{\widehat{H}=1}{\underset{\widehat{H}=0}{\gtrless}%
}\upsilon 
\end{equation}
where $\upsilon$ denotes an appropriate threshold derived  for example from a  Bayes' criteria that minimizes the average probability of error $P_e$.

To be specific, we consider the simple binary hypothesis problem
\begin{equation}
\label{ynstatistics}
H_{m}:\:\:y_{n}=\mu_{m}+\xi_n,\:\xi_n \sim
        \mathcal{N}\left(0,\sigma^{2}\right), \: m=0,1
\end{equation}
where, without loss of generality, we let
$\mu_{1}=-\mu_{0}=\mu$. 

\emph{Parallel architecture: fusion center.} Under this model, the local
likelihoods $r_{n}$ are also Gaussian, i.e.,
\begin{equation}
\label{rnstatistics}
H_{m}:\:\: r_{n} \sim \mathcal{N}\left(\frac{2\mu
        \mu_{m}}{\sigma^{2}},\frac{4\mu^{2}}{\sigma^{2}}\right)
\end{equation}
From~(\ref{centraldecision}), the test statistic for the parallel architecture fusion center is also Gauss
\begin{equation}
\label{ellstatistics}
H_{m}:\:\: \ell\sim{\cal N}\left(\frac{2\mu
        \mu_{m}}{\sigma^{2}},\frac{4\mu^{2}}{N\sigma^{2}}\right)
\end{equation}
The error performance of the minimum probability of error $P_e$ Bayes' detector (threshold $\upsilon=0$ in~(\ref{centraldecision})) is given by
\begin{equation}
\label{centralizedPe}
P_e=\mbox{erfc}^{\star}\left(\frac{d}{2}\right)=\int_{d/2}^{+\infty}\frac{1}{\sqrt{2\pi}}e^{-\frac{x^2}{2}}\,dx
\end{equation}
where the equivalent signal to noise ratio~$d^2$ that characterizes the performance is given by, \cite{VanTreesPartI},
\begin{equation}
\label{equivalentSNRd}
d=\frac{2\mu\sqrt{N}}{\sigma}
\end{equation}

\emph{Distributed detection.}
We now consider a distributed solution where the  sensor nodes reach a global common decision~$\widehat{H}$ about the true state~$H$ based on the measurements collected by all sensors but through \textit{local
exchange} only of information over the network~$G$. By \textit{local
exchange}, we mean that the sensor nodes do not have the ability to
\textit{route} their data to parts of the network other than their
immediate neighbors. Such algorithms are of course of practical significance when using power and complexity constrained sensor nodes since such sensor networks may not be able to handle the high costs associated with routing or flooding techniques. We apply the average-consensus algorithm described in Section~\ref{sub:consensusalgorithm}.
This distributed average-consensus detector achieves asymptotically (in the number of iterations) the same optimal error performance $P_e$ of the parallel architecture given by~(\ref{centralizedPe}), see~\cite{SensNets:AldosariAsilomar05,aldosarimouramay06}. 

Actually, we consider a more general problem than the average-consensus algorithm in~(\ref{Eqn:Update x0}), namely, we assume that the communications among sensors is through noisy channels. Let the network state, i.e., the likelihood vector, at iteration~$i$ be $\mathbf{x}_{i}\in \mathbb{R}^{N}$. We modify~(\ref{Eqn:Update x0}), by taking into account the communication channel noise in each iteration. The distributed detection average-consensus algorithm is modeled by
\begin{equation}
\label{Fusion_Model} 
\mathbf{x}_{i+1}=W\mathbf{x}_{i} +
\mathbf{n}_{i}
\end{equation}
The weight matrix is as given by~(\ref{iter}) using the weight in~(\ref{Eqn: opt const weights})
\begin{equation}
\label{optimalweightmatrix}
W = I-\frac{2}{\lambda_{2}(L)+\lambda_{N}(L)}L
\end{equation}
The initial condition~$\mathbf{x}_0$ that collects the local LLRs~$r_n$ given 
in~(\ref{xinitial}), herein repeated,
\[
\mathbf{x}_0=\left[r_1\cdots r_N\right]^T
\]
has statistics
\begin{equation}
\label{init_lik_dist} 
H_{m}:\,\,\,\mathbf{x}_{0}\sim\mathcal{N}\left(\frac{2\mu\mu_{m}}{\sigma^{2}}\mathbf{1},\Sigma_0=\frac{4\mu^{2}}{\sigma^{2}}I\right),\,\,\,
m=0,1
\end{equation}
 The communications noise at iteration~$i$ is zero mean Gauss white noise with covariance~$R$ given by
\begin{eqnarray}
\label{nistatistics}
\mathbf{n}_{i}&\sim& \mathcal{N}\left(\mathbf{0},R\right)\\
\label{Rcovariance}
R&=&\mbox{diag}\left[\phi_{1}^{2},...,\phi_{N}^{2}\right]
\end{eqnarray}
The \emph{communication channel} noise $\mathbf{n}_i$ is assumed to be independent of the \emph{measurement} noise $\xi_n$, $\forall i,n$.

The final decision at each sensor is 
\[
x_{n}(i)\overset{\widehat{H}(n)=1}{\underset{\widehat{H}(n)=0}{\gtrless}%
}\upsilon
\]
where $\widehat{H}(n)$ denotes the decision of sensor $v_{n}$.
\subsection{Noise Analysis}
\label{noiseanalysis}
In this Subsection we carry out the statistical analysis of the distributed average-consensus detector. 
\begin{theorem}
\label{thm:mean}
The local state~$x_n(i)$ has mean
\begin{equation}
\label{meanxni}
H_m:\,\,\,\,\, E\left[x_n(i)\right]=\frac{2\mu\mu_m}{\sigma^2}
\end{equation}
where $E[\cdot]$ stands for the expectation operator and $\mu_m$ is either $\mu_1=\mu$ or $\mu_0=-\mu$.
\end{theorem}
\begin{proof}
From the distributed detection~(\ref{Fusion_Model})
\begin{equation}
x_{n}(i)=\sum_{j=1}^{N}\left(W^{i}\right)_{n,j}r_j
\end{equation}
Hence,
\begin{equation}
\label{like_dist} 
E\left[x_{n}(i)\right]=\frac{2\mu\mu_{m}}{\sigma^{2}}\sum_{j=1}^{N}\left(W^{i}\right)_{n,j}
\end{equation}
It follows:
\begin{eqnarray}
\sum_{j=1}^{N}\left(W^{i}\right)_{n,j} & = & \left(W^{i}\mathbf{1}\right)_{n,1} \nonumber \\
                          & = & \mathbf{1}_{n,1} \nonumber \\
                          & = & 1
\end{eqnarray}
(since $\mathbf{1}$ is an eigenvector of $W$ with eigenvalue $1$,
it is also an eigenvector of $W^{i}$ with eigenvalue $1$.) Replacing this result in~(\ref{like_dist}) leads to the Theorem and~(\ref{meanxni}).
\end{proof}
We now consider the variance $\mbox{var}_n(i)$ of the state $x_n(i)$ of the sensor~$n$ at iteration~$i$. The following Theorem provides an upper bound.
\begin{theorem}
\label{thm:varbound}
The variance $\mbox{var}_n(i)$ of the state $x_n(i)$ of the sensor~$n$ at iteration~$i$ is bounded by
\begin{equation}
\label{eq:varbound}
\mbox{var}_n(i)\leq 
\frac{4\mu^{2}}{\sigma^{2}}\left[\frac{1}{N}+\gamma_{2}^{2i}\left(1-\frac{1}{N}\right)\right] +
\phi_{\mbox{\scriptsize max}}^{2}\left[\frac{i}{N} +
\frac{1-\gamma_{2}^{2i}}{1-\gamma_{2}^{2}}\left(1-\frac{1}{N}\right)\right]
\end{equation}
where $\gamma_2$ is given in~(\ref{gamma2-2}).
\end{theorem}
\begin{proof}
 Let the covariance of the network state at iteration~$i$ be
\[
\Sigma_i=\mbox{covar}\{\mathbf{x}_i\}
\] 
From eqn.~(\ref{Fusion_Model}) and using standard stochastic processes analysis
\begin{equation}
\label{cov_mat} \Sigma_{i} = W^{i}\Sigma_{0}W^{i} +
\sum_{k=0}^{i-1}W^{k}RW^{k}
\end{equation}
Thus the variance at the $n$-th sensor is given by,
\begin{equation}
\label{var_p_i} 
\mbox{var}_{n}(i) =
\left(W^{i}\Sigma_{0}W^{i}\right)_{n,n} + \sum_{k=0}^{i-1}\left(W^{k}RW^{k}\right)_{n,n}
\end{equation}
Let $\mathbf{w}_{j}^{(k)}$ be the columns of $W^k$, $j\in [1,...,N]$.
Then,
\begin{equation}
\label{WRW_dec} 
W^{k}RW^{k} =
\sum_{j=1}^{N}\phi_{j}^{2}\mathbf{w}_{j}^{(k)}\mathbf{w}_{j}^{(k)^{T}}
\end{equation}
It follows
\begin{equation}
\label{WRWpp} 
\left(W^{k}RW^{k}\right)_{n,n} =
\sum_{j=1}^{N}\phi_{j}^{2}\left(w_{j,n}^{(k)}\right)^{2}
\end{equation}
where $w_{j,n}^{(k)}$ represents the $n$-th component of the vector
$\mathbf{w}_{j}^{(k)}$.
Denote by
\[
\phi_{\mbox{\scriptsize max}}=\max\left(\phi_{1},...,\phi_{N}\right)
\]
From eqn.~(\ref{WRWpp}), we get
\begin{eqnarray}
\label{WRWpp_bound} 
\left(W^{k}RW^{k}\right)_{n,n} & \leq &
\phi_{\mbox{\scriptsize max}}^{2}\sum_{j=1}^{N}\left(w_{j,n}^{(k)}\right)^{2} \nonumber \\
& = & \phi_{\mbox{\scriptsize max}}^{2}\left(W^{2k}\right)_{n,n}
\end{eqnarray}
We now use the eigendecomposition of~$W$ in~(\ref{eq:Wexpansion}). 
 This leads to
\begin{equation}
\label{Wk_dec} 
W^{2k} =
\sum_{m=1}^{N}\gamma_{m}^{2k}\mathbf{u}_{m}\mathbf{u}_{m}^{T}
\end{equation}
from which
\begin{eqnarray}
\label{W2kpp_bound} 
\left(W^{2k}\right)_{n,n} & = &
\sum_{m=1}^{N}\gamma_{m}^{2k}\left(\mathbf{u}_{m,n}\right)^{2} \nonumber \\
& = & \frac{1}{N} + \sum_{m=2}^{N}\gamma_{m}^{2k}(\mathbf{u}_{m,n})^{2} \nonumber \\
& \leq & \frac{1}{N} +\gamma_{2}^{2k}\sum_{m=2}^{N}\left(\mathbf{u}_{m,n}\right)^{2}
\nonumber
\\ & = & \frac{1}{N} + \gamma_{2}^{2k}\left(1-\frac{1}{N}\right)
\end{eqnarray}
Hence, from eqn.~(\ref{WRWpp_bound}),
\begin{eqnarray}
\label{WRWpp_bound_1} 
\left(W^{k}RW^{k}\right)_{n,n} & \leq &
\phi_{\mbox{\scriptsize max}}^{2}\left(W^{2k}\right)_{n,n} \nonumber \\
& \leq &
\phi_{\mbox{\scriptsize max}}^{2}\left[\frac{1}{N} + \gamma_{2}^{2k}\left(1-\frac{1}{N}\right)\right]
\end{eqnarray}
Through a similar set of manipulations,
\begin{eqnarray}
\label{WSWpp_bound} \left(W^{i}\Sigma_{0}W^{i}\right)_{n,n} & = &
\frac{4\mu^{2}}{\sigma^{2}}\left(W^{2i}\right)_{n,n} \nonumber \\ 
& \leq &
\frac{4\mu^{2}}{\sigma^{2}}\left[\frac{1}{N}+\gamma_{2}^{2i}\left(1-\frac{1}{N}\right)\right]
\end{eqnarray}
Finally from eqn.~(\ref{var_p_i}) we obtain,
\begin{eqnarray}
\label{var_p_i_bound} 
\mbox{var}_{n}(i) & \leq &
\frac{4\mu^{2}}{\sigma^{2}}\left[\frac{1}{N}+\gamma_{2}^{2i}\left(1-\frac{1}{N}\right)\right] +
\sum_{k=0}^{i-1}\phi_{\mbox{\scriptsize max}}^{2}\left[\frac{1}{N} +
\gamma_{2}^{2k}\left(1-\frac{1}{N}\right)\right] \nonumber \\ & = &
\frac{4\mu^{2}}{\sigma^{2}}\left[\frac{1}{N}+\gamma_{2}^{2i}\left(1-\frac{1}{N}\right)\right] +
\phi_{\mbox{\scriptsize max}}^{2}\sum_{k=0}^{i-1}\left[\frac{1}{N} +
\gamma_{2}^{2k}\left(1-\frac{1}{N}\right)\right] \nonumber \\ 
& = &
\frac{4\mu^{2}}{\sigma^{2}}\left[\frac{1}{N}+\gamma_{2}^{2i}\left(1-\frac{1}{N}\right)\right] +
\phi_{\mbox{\scriptsize max}}^{2}\left[\frac{i}{N} +
\frac{1-\gamma_{2}^{2i}}{1-\gamma_{2}^{2}}\left(1-\frac{1}{N}\right)\right]
\end{eqnarray}
which gives an upper bound on the variance of the $n$-th sensor at
iteration~$i$ and proves Theorem~\ref{thm:varbound}.
\end{proof}
If the channels are noiseless, we immediately obtain a Corollary to Theorem~\ref{thm:varbound} that bounds the variance of the state of sensor~$n$ at iteration~$i$.
\begin{corollary}
\label{cor:varbound}
With noiseless communication channels, the variance of the state of sensor~$n$ at iteration~$i$ is bounded by
\begin{equation}
\label{eqn:corvarbound}
\mbox{var}_n(i)\leq
\frac{4\mu^{2}}{\sigma^{2}}\left[\frac{1}{N}+\gamma_{2}^{2i}\left(1-\frac{1}{N}\right)\right] 
\end{equation}
\end{corollary}
We now interpret Theorems~\ref{thm:mean} and~\ref{thm:varbound}, and Corollary~\ref{cor:varbound}. Theorem~\ref{thm:mean} shows that the mean of the local state is the same as the mean of the global statistic~$\ell$ of the fusion center in the parallel architecture. Then to compare the local probability of error $P_e(i,n)$ at sensor~$n$ and iteration~$i$ in the distributed detector with the probability of error $P_e$ of the fusion center in the parallel architecture we need to compare the variances of the sufficient statistics in each detector.  With noiseless communication channels, we see that the upper bound in~(\ref{eqn:corvarbound}) in Corollary~\ref{cor:varbound} converges to
We now interpret Theorems~\ref{thm:mean} and~\ref{thm:varbound}, and Corollary~\ref{cor:varbound}. Theorem~\ref{thm:mean} shows that the mean of the local state is the same as the mean of the global statistic~$\ell$ of the fusion center in the parallel architecture. Then to compare the local probability of error $P_e(i,n)$ at sensor~$n$ and iteration~$i$ in the distributed detector with the probability of error $P_e$ of the fusion center in the parallel architecture we need to compare the variances of the sufficient statistics in each detector.  With noiseless communication channels, we see that the upper bound in~(\ref{eqn:corvarbound}) in Corollary~\ref{cor:varbound} converges to
\[
\frac{4\mu^{2}}{\sigma^{2}}\left[\frac{1}{N}+\gamma_{2}^{2i}\left(1-\frac{1}{N}\right)\right] 
\rightarrow \frac{4\mu^{2}}{N\sigma^{2}}
\]
which is the variance of the parallel architecture test statistic~(\ref{centraldecision}). This shows that
\begin{equation}
\label{convergencepeitope}
\lim_{i\rightarrow \infty} P_e(i,n)=P_e
\end{equation}
The rate of convergence is again controlled  by
\[
\gamma_{2}^{2i}=\left(1-\frac{2\lambda_2(L)}{\lambda_2(L)+\lambda_N(L)}\right)^{2i}
\]
and maximizing this rate is equivalent to minimizing $\gamma_2$, which in turn, see~(\ref{gamma2-2}), is equivalent to maximizing the eigenratio parameter $\gamma=\lambda_2(L)/\lambda_N(L)$ like for the average-consensus algorithm.

For noisy channels, it is interesting to note that there is a linear trend $\phi_{\mbox{\scriptsize max}}^{2}i/N$ that makes $\mbox{var}_n(i)$ to become arbitrarily large as the number of iterations~$i$ grows to~$\infty$. We no longer have the convergence of the probability of error $P_e(i,n)$ as in~(\ref{convergencepeitope}).
The average minimum probability of error is still given by~(\ref{centralizedPe}), with now the equivalent SNR parameter~$d^2$ bounded below by Theorem~\ref{thm:varbound}.

\subsection{Optimal number of iterations}
\label{optimalnumberiterations}
With noisy communication channels, the performance of the distributed detector no longer achieves the performance of the fusion center in a parallel architecture. This is no surprise, since each iteration corrupts the inter communicated state of the sensor. However, there is an interesting tradeoff between sensing signal to noise ratio~(S-SNR) and the communication noise. Intuitively, the local sensors perceive better the global state of the environment as they obtain information through their neighbors  from more remote sensors. However, this new information is counter balanced by the additional noise introduced by the communication links. This leads to an interesting tradeoff that we now exploit and leads to an optimal number of iterations to carry out the consensus through noisy channels before a decision is declared by each sensor.

 The upper bound in eqn.~(\ref{var_p_i_bound})
is a function of the number of iterations~$i$. We rewrite it,  replacing the integer valued iteration number~$i$ by a continuous variable~$z$, as
\begin{equation}
\label{f_def} \emph{f}(z) =
\left(\frac{4\mu^{2}}{N\sigma^{2}}+\frac{\phi_{\mbox{\scriptsize max}}^{2}\left(1-\frac{1}{N}\right)}{1-\gamma_{2}^{2}}\right)
+
\left(1-\frac{1}{N}\right)\left(\frac{4\mu^{2}}{\sigma^{2}}-\frac{\phi_{\mbox{\scriptsize max}}^{2}}{1-\gamma_{2}^{2}}\right)\gamma_{2}^{2z}
+ \frac{\phi_{\mbox{\scriptsize max}}^{2}}{N}z
\end{equation}
We consider only the case when
\begin{equation}
\label{power_assumption} \frac{4\mu^{2}}{\sigma^{2}} >
\frac{\phi_{\mbox{\scriptsize max}}^{2}}{1-\gamma_{2}^{2}}
\end{equation}
This is  reasonable. For example, if $\frac{4\mu^{2}}{\sigma^{2}} >
\phi_{\mbox{\scriptsize max}}^{2}$, which is the case when the communication noise is smaller than the equivalent sensing noise power and iterating among sensors can be reasonably expected to improve upon decisions based solely on the local measurement. Secondly, if $\gamma_2$, which is bounded above by~$1$, is small, then the right-hand-side of~(\ref{power_assumption}) is more likely to be satisfied. This means that topologies like the Ramanujan graphs where $\gamma_2$ is minimized (which, from~(\ref{gamma2-2}) means that the eigenratio parameter~$\gamma$ is maximized) will satisfy better this assumption.

We now state the result on the number of iterations.
\begin{theorem}
\label{thm:noiterations}
If~(\ref{power_assumption}) holds, $\emph{f}(z)$ has a global minimum at
\begin{equation}
\label{z_ast} 
z^{\ast} = \frac{1}{2\ln\, \gamma_{2}}\ln\,
\left(\frac{\phi_{\mbox{\scriptsize max}}^{2}}{\left(2\ln\,
\frac{1}{\gamma_{2}}\right)(N-1)\left(\frac{4\mu^{2}}{\sigma^{2}}-\frac{\phi_{\mbox{\scriptsize max}}^{2}}{1-\gamma_{2}^{2}}\right)}\right)
\end{equation}
\end{theorem}
\begin{proof}
When~(\ref{power_assumption}) holds, $\emph{f}(z)$ is
convex. Hence there exists a global minimum, say 
attained at $z^{\ast}$. We
find $z^{\ast}$ by rooting the first derivative, successively obtaining
\begin{eqnarray}
\label{cond_min_f} 
\frac{d\emph{f}}{dz}\left(z^{\ast}\right)  &=& \left(2\ln\,
\gamma_{2}\right)\left(1-\frac{1}{N}\right)\left(\frac{4\mu^{2}}{\sigma^{2}}-\frac{\phi_{\mbox{\scriptsize max}}^{2}}{1-\gamma_{2}^{2}}\right)\gamma_{2}^{2z^{\ast}}
+ \frac{\phi_{\mbox{\scriptsize max}}^{2}}{N} = 0\\
\label{min_f_z} 
\gamma_{2}^{2z^{\ast}} &=&
-\frac{\phi_{\mbox{\scriptsize max}}^{2}}{(N-1)\left(2\ln\,
\gamma_{2}\right)\left(\frac{4\mu^{2}}{\sigma^{2}}-\frac{\phi_{\mbox{\scriptsize max}}^{2}}{1-\gamma_{2}^{2}}\right)}\\
\label{z_ast-1} 
z^{\ast} &=& \frac{1}{2\ln\, \gamma_{2}}\ln\,
\left(\frac{\phi_{\mbox{\scriptsize max}}^{2}}{\left(2\ln\,
\frac{1}{\gamma_{2}}\right)(N-1)\left(\frac{4\mu^{2}}{\sigma^{2}}-\frac{\phi_{\mbox{\scriptsize max}}^{2}}{1-\gamma_{2}^{2}}\right)}\right)
\end{eqnarray}
\end{proof}
From Theorem~\ref{thm:noiterations}, we conclude that, if $z^{\ast}>0$, then the variance upper bound will decrease till $i^{\ast} = \lfloor z^{\ast}\rfloor $. The iterative distributed detection algorithm should be continued till $i^{\ast}$ if 
\begin{equation}
\label{maxiterationnumber}
\min\left(\emph{f}\left(\lfloor z^{\ast}\rfloor\right),\emph{f}\left(\lceil
z^{\ast}\rceil\right)\right) < \mbox{var}_{n}(0)=\frac{4\mu^{2}}{\sigma^{2}}
\end{equation}

{\bf Numerical Examples.} We illustrate Theorem~\ref{thm:noiterations} with two numerical examples. We consider a network of $N=1,000$ sensors, $\mu^{2}/\sigma^{2} = 1$ (0 db), and $\gamma_{2} = .7$. The initial likelihood variance before fusion is $\mbox{var}_{n}(0) = 4$. We first consider
$\phi_{\mbox{\scriptsize max}} = .1$
Then, $z^{\ast} = 17.6$  and $\mbox{var}_{n}(17)\leq \emph{f}(\lfloor
z^{\ast} \rfloor) = .0238=\emph{f}(\lceil z^{\ast} \rceil)$.  The variance reduction achieved with iterative distributed detection over the single measurement decision is $\frac{\mbox{var}_{n}(0)}{\mbox{var}_{n}(i^{\ast})}
\geq 168=22~\mbox{dB}$, a considerable improvement.
We now consider a second case where the communication noise is 
$\phi_{\mbox{\scriptsize max}}=.3162$. It follows that $z^{\ast}=14.3$, and the improvement by iterating till $i^{\ast}=14$ with the distributed detection is 
$\frac{\mbox{var}_{n}(0)}{\mbox{var}_{n}(14)}\geq 20=13~\mbox{dB}$.

\section{Experimental Results}
\label{results}
This section shows how Ramanujan graph topologies outperform other topologies. We first describe the graph topologies to be contrasted with the Ramanujan LPS-II constructions described in Section~\ref{ramanujan}. 
We start by defining the average degree
$k_{\mbox{\scriptsize{avg}}}$ of a graph~$G$ as
\[
k_{\mbox{\scriptsize{avg}}}=\frac{2|E|}{|V|}
\]
where $|E|=M$ denotes the number of edges and $|V| = N$ is the number of
vertices of the graph. In this section, we use the symbols and terms $k$ and
$k_{\mbox{\scriptsize{avg}}}$ interchangeably. For, $k$-regular
graphs, it follows that $k_{\mbox{\scriptsize{avg}}}=k$. This
means, that, when we work with general graphs, $k$ refers to the
average degree, while with regular graphs, it refers to both the
average degree and the degree of each vertex.

\subsection{Structured graphs, Watts-Strogatz Graphs, and Erd\"os-Ren\'{y}i Graphs}
\label{othergraphs}
We compare Ramanujan graphs, which are regular graphs, with regular and non regular graphs. The symbol~$k$ will stand in this Section for the degree of the graph for regular graphs and for the average degree for non regular graphs. We describe briefly the three classes of graphs used to benchmark the Ramanujan graphs. Structured graphs  usually have high clustering but large average path length. Erd\"os-Ren\'{y}i graphs are random graphs, they have small average path length but low clustering. Small-world graphs generated with a rewiring probability above a phase transition threshold have both high clustering and  small average path length.

{\bf Structured graphs: Regular ring lattice (RRL.} This is a highly structured network.  The nodes are numbered sequentially (for simplicity, display them uniformly placed on a ring.) Starting from node~\#~1, connect each node to $k/2$ nodes to the left and $k/2$ nodes to the right. The resulting graph is regular with degree~$k$.

{\bf Small world networks: Watts-Strogatz (WS-I).} We  explain briefly the Watts-Strogatz
construction of a small world network, \cite{SensNets:Watts98}. It
starts from a  highly structured regular network where the nodes
are placed uniformly around a circle, with each node connected to
its $k$ nearest neighbors. Then, random rewiring is conducted on
all graph links. With probability $p_{w}$, a link is rewired to a
different node chosen uniformly at random. Notice that the $p_{w}$
parameter controls the ``randomness'' of the graph in the sense
that $p_{w}=0$ corresponds to the original highly structured
network while $p_{w}=1$ results in a random network. Self and
parallel links are prevented in the rewiring procedure and the
number of links is kept constant, regardless of the value of
$p_{w}$. In~\cite{SensNets:AldosariAsilomar05}, distributed
detection was studied with two slightly different versions of the
Watts-Strogatz model. In both versions, the rewiring procedure is
such that the nodes are considered one by one in a fixed direction
along the circle (clockwise or counter clockwise.) For each node,
the $k/2$ edges connecting it to the following nodes (in the same
direction) are rewired with probability $p_{w}$. In the first
version of the Watts-Strogatz model, called Watts-Strogatz-I
(WS-I) in the sequel, the edges are kept connected to the current
node while their other ends are rewired with probability $p_{w}$.
In the second version, called Watts-Strogatz-II (WS-II), the
particular vertex to be disconnected is chosen randomly between
the two ends of the rewired edges. It was shown in~\cite{SensNets:AldosariAsilomar05} that the WS-I graphs yield
better convergence rates among the different models of small world
graphs considered in that paper (WS-I, WS-II, and the Kleinberg
model, \cite{SensNets:KleinbergNature2000,SensNets:KleinbergThComp2000}.)  Hence, we restrict attention here to WS-I graphs.

{\bf Erd\"os-Ren\'{y}i random graphs (ER).}  In these graphs, we randomly choose $\frac{Nk}{2}$ edges out of a total of $\frac{N(N-1)}{2}$ possible edges. These are not regular graphs, their degree distribution follows a binomial distribution, which in the limit of large~$N$ approaches the Poisson law.

\subsection{Comparison Studies}
\label{comparisonstudies}
We present numerical studies that will show the superiority of the Ramanujan graphs~(RG) over the other three classes of graphs: Regular ring lattice~(RRL), Watts-Strogatz-I (WS-I), and Erd\"os-Ren\'{y}i~(ER) graphs.  We carry out three types of comparisons:
\begin{inparaenum}[(1)]
\item Convergence speed $S_{\mbox{\scriptsize{c}}}$;
\item The $\gamma$ parameters for the RG and each of the other three classes of graphs;
\item The algebraic connectivity $\lambda_2(L)$ for the RG and each of the other three classes of graphs. 
\end{inparaenum}
In Section~\ref{distributedinference}, we considered a distributed detection problem based on the average-consensus algorithm. Here we present results for  the noiseless link case. We define the convergence time
$T_{\mbox{\scriptsize{c}}}$ of the distributed detector, as the number of iterations required
to reach within $10 \%$ of the global probability of error,
averaged over all sensor nodes. Rather than using~$T_{\mbox{\scriptsize{c}}}$, the results are presented in terms of the convergence speed, $S_{\mbox{\scriptsize{c}}}=1/T_{\mbox{\scriptsize{c}}}$.  
To simplify the comparisons, we subscript the $\gamma$ parameter by the corresponding acronym, e.g., $\gamma_{\mbox{\scriptsize RG}}$ to represent the eigenratio of the Ramanujan graph. We also define the following comparison parameters 
\begin{equation}
\psi(\mbox{RRL})=\frac{S_{\mbox{\scriptsize c, RG}}}{S_{\mbox{\scriptsize c, RRL}}},\hspace{.25cm}
\nu(\mbox{RRL})=\frac{\gamma_{\mbox{\scriptsize RG}}}{\gamma_{\mbox{\scriptsize RRL}}},\hspace{.25cm}\mbox{and}\hspace{.25cm}
\eta(\mbox{RRL})=\frac{\lambda_{2,\mbox{\scriptsize RG}}(L)}{\lambda_{2,\mbox{\scriptsize RRL}}(L)}
\end{equation}

\begin{figure}[htb]
\begin{center}
\includegraphics[height=2.1in, width=2.1in ]{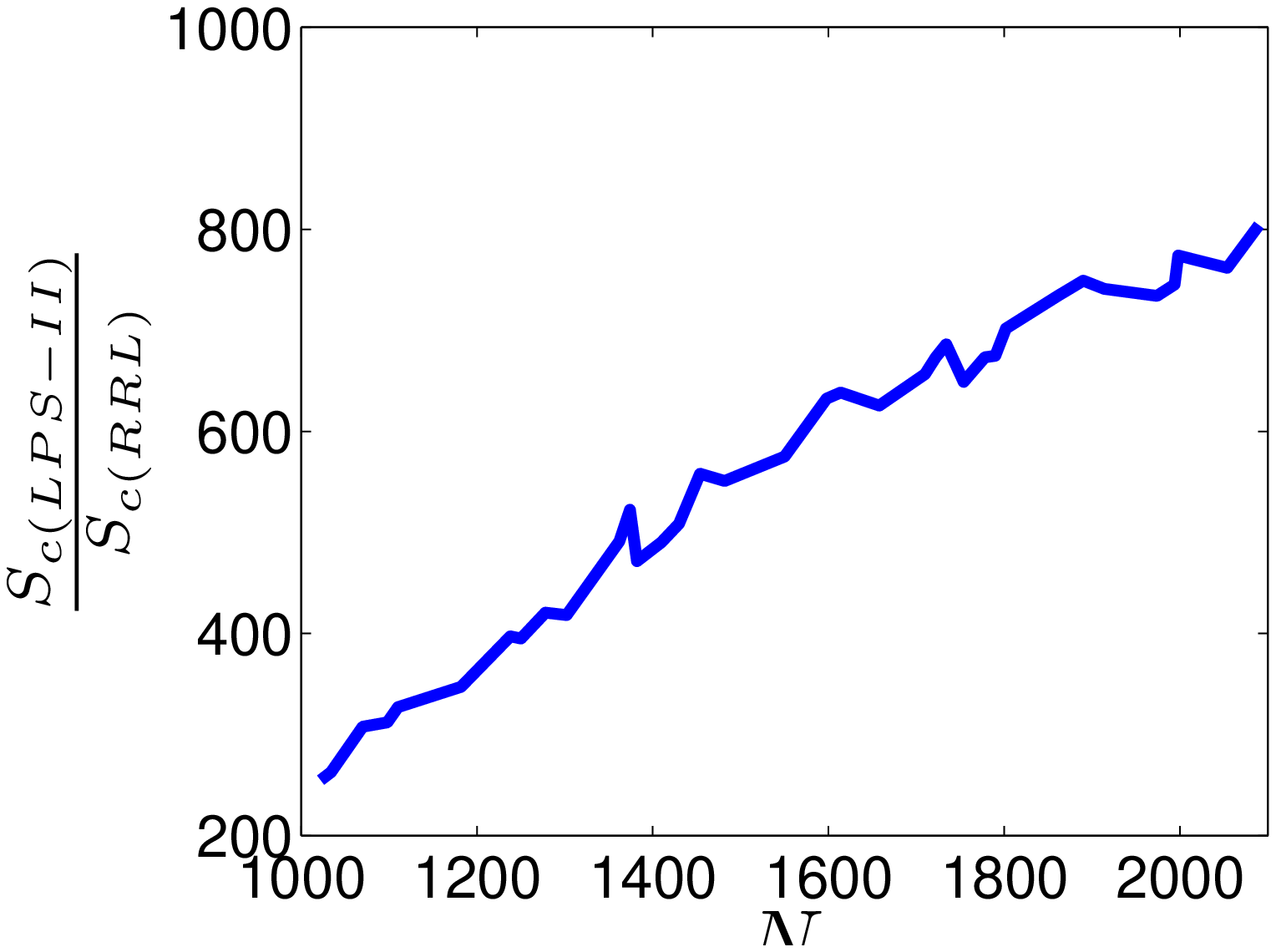}
\includegraphics[height=2.1in, width=2.1in ] {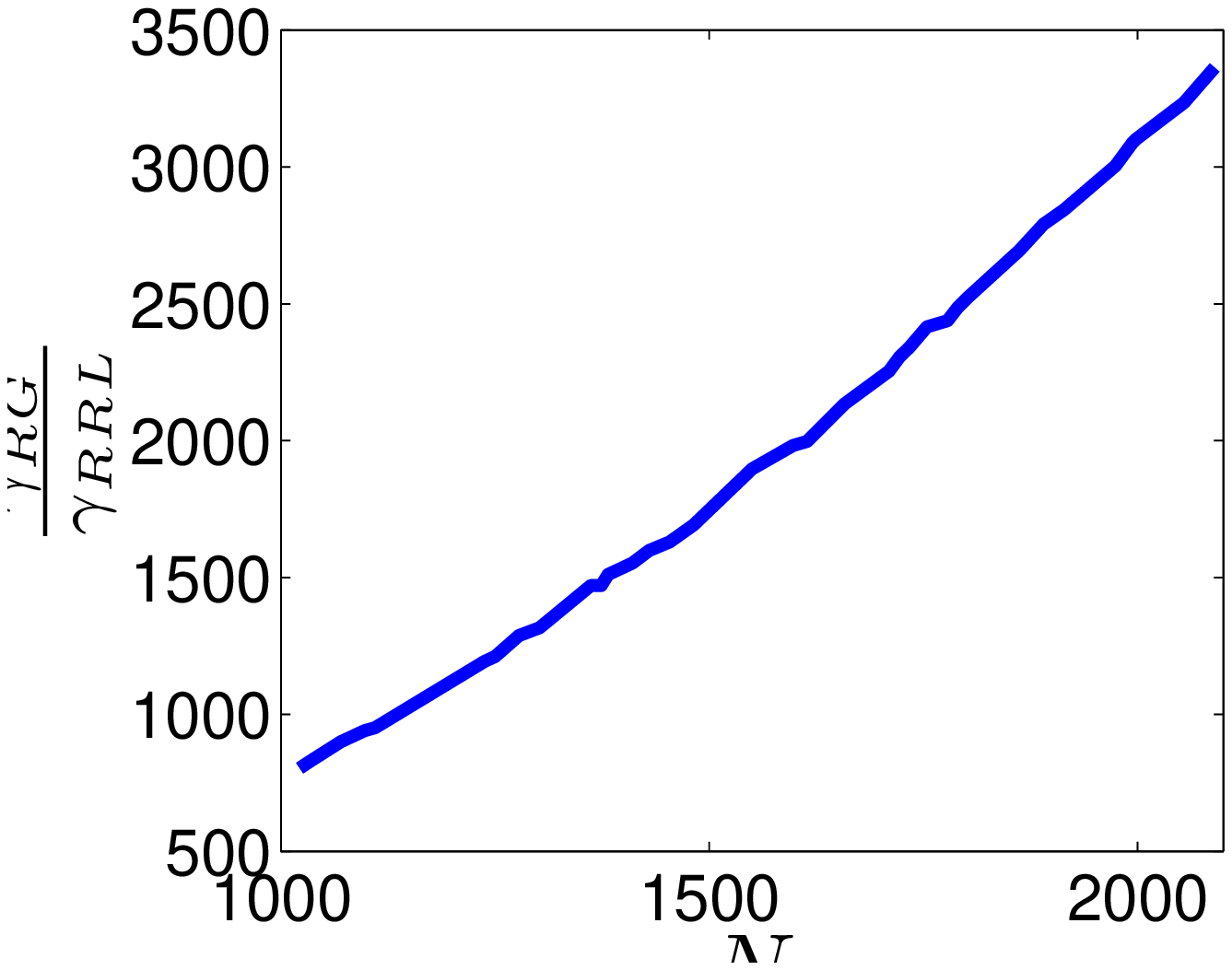}
\includegraphics[height=2.1in, width=2.1in ] {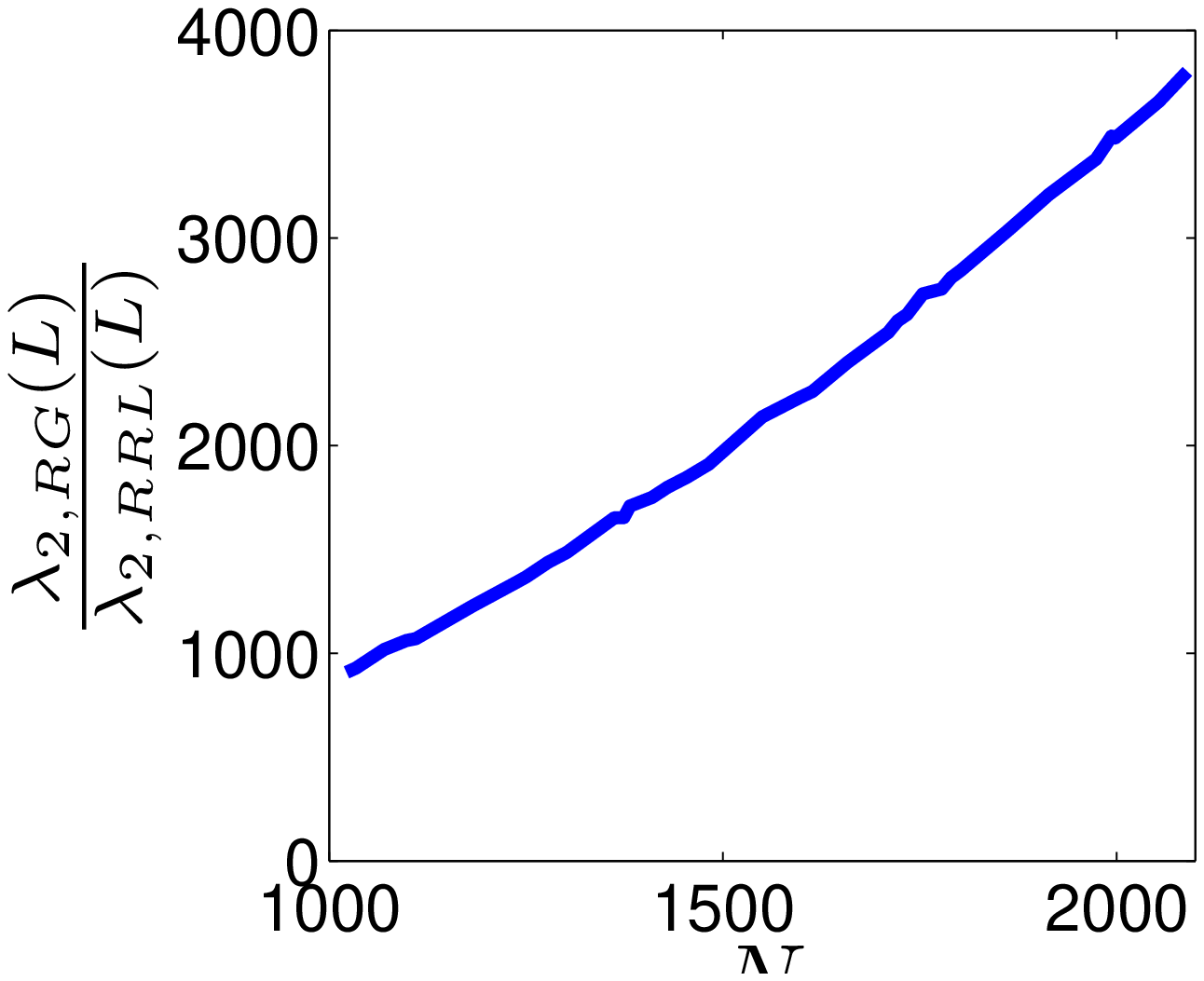}
\caption{Spectral properties of LPS-II and
RRL graphs, $k=18$, varying~$N$: Left: Ratio of convergence speed~$\psi(\mbox{RRL})$; Center: Ratio~$\nu(\mbox{RRL})$ of $\frac{\lambda_{2}(L)}{\lambda_{N}(L)}$;
Right: Ratio~$\eta(\mbox{RRL})$ of $\lambda_{2}(L)$.}
\label{speed_ratio_lattice}
\label{speed_ratio_lattice-2}\label{lambda_ratio_ringvsram} \label{lambda_2_ringvsram}
\end{center}
\end{figure}

{\bf Ramanujan graphs and regular ring lattices.}
Fig.~\ref{speed_ratio_lattice}  compares RG with RRL graphs. The panel on the right plots~$\psi(\mbox{RRL})$, 
the  center panel displays~$\nu(\mbox{RRL})$, and the  right panel shows~$\eta(\mbox{RRL})$   when the degree~$k=18$ and the number of nodes~$N$ varies. We conclude that  the~RGs converge $3$~orders of magnitude faster than the~RRLs,  the $\gamma$ parameters can be up to $3,500$ times faster, and  the algebraic connectivity for the RGs can be up to $4,000$ times larger than for the RRLs.

\begin{figure}
\begin{center}
\includegraphics[height=2in, width=2in ] {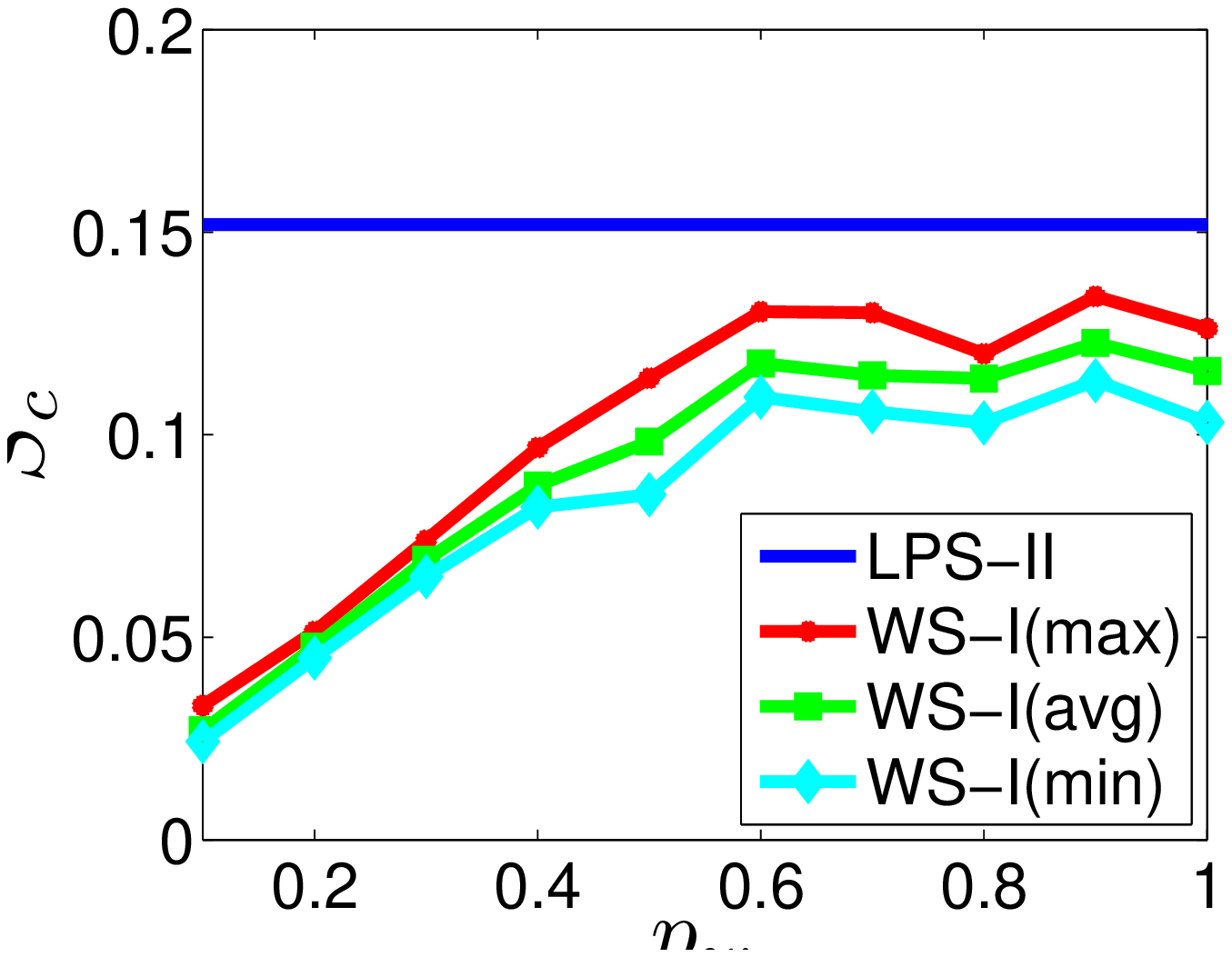}
\includegraphics[height=2in, width=2in ] {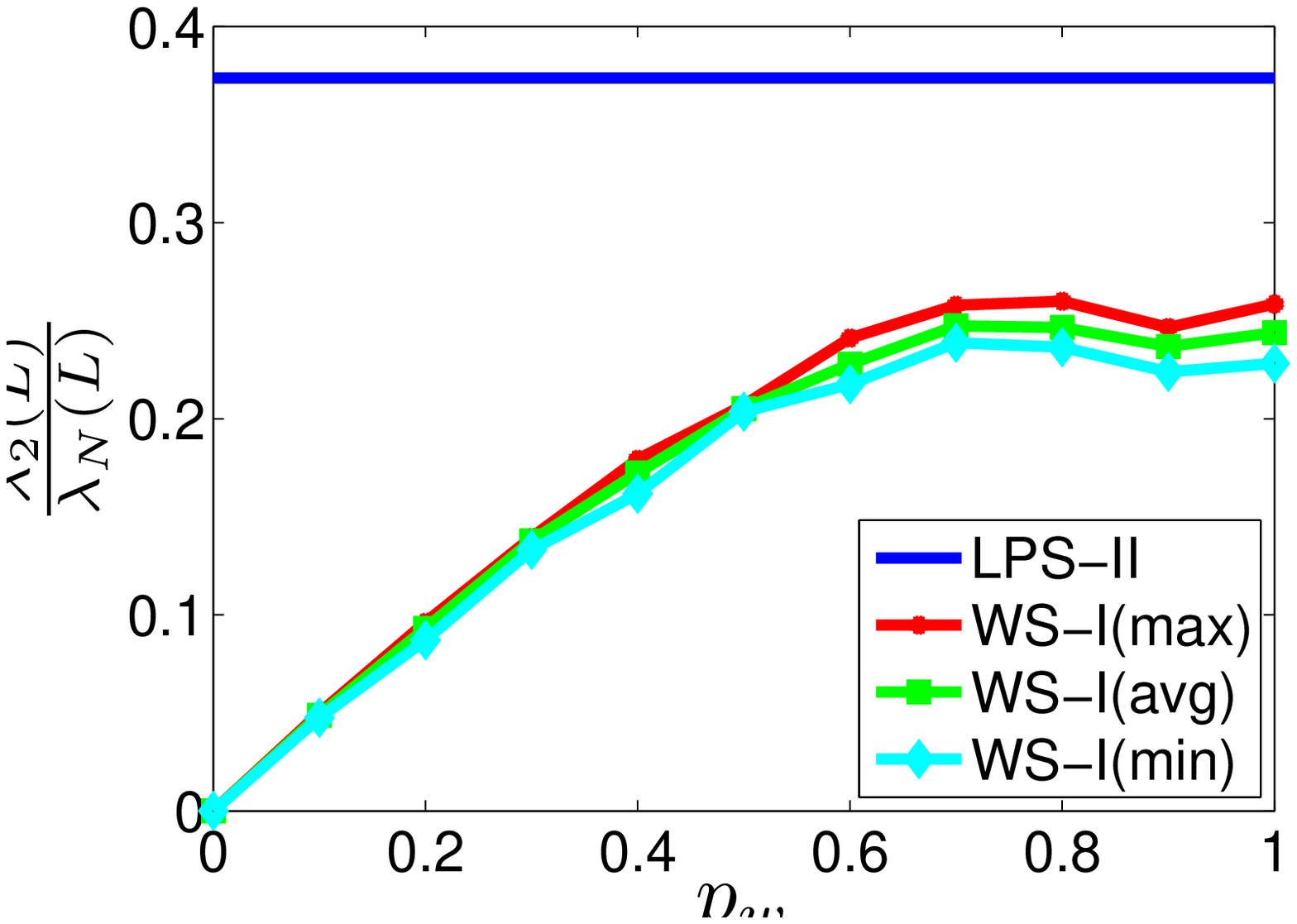}
\includegraphics[height=2in, width=2in ] {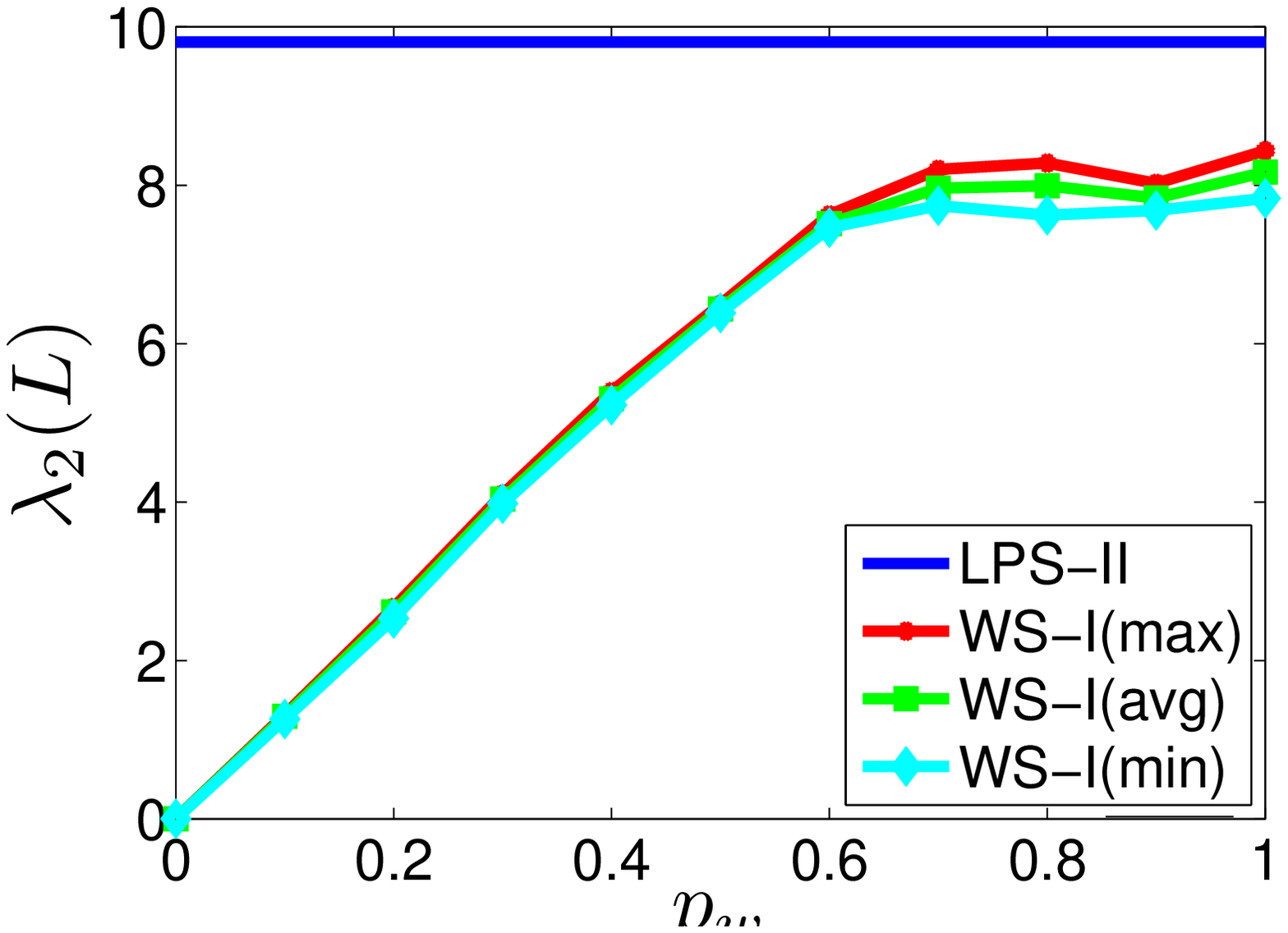}
\caption{Spectral properties of LPS-II and
WS-I graphs, $N=6038$, $k=18$, varying~$p_{\mbox{\scriptsize{w}}}$
Left: $S_{\mbox{\scriptsize{c}}}$;
Center: eigenratio $\gamma=\frac{\lambda_{2}(L)}{\lambda_{N}(L)}$; Right: 
algebraic connectivity $\lambda_{2}$.}
\label{eig_WSI_6038_ratio_mod} 
\label{eig_WSI_6038_ratio_mod-2} \label{eig_WSI_6038_lambda_2}\label{Sc_WSI_6038}
\end{center}
\end{figure}

{\bf Ramanujan graphs and Watts-Strogatz graphs.}
Fig.~\ref{Sc_WSI_6038}  contrasts the RG with the WS-I graphs. Because the WS-I graphs are randomly generated, we fix the  number of nodes $N=6038$ and the degree $k=18$ and vary on the horizontal axis the rewiring probability $0\leq p_{\mbox{\scriptsize{w}}}\leq 1$.  The Figure shows on the left panel the convergence speed $S_{\mbox{\scriptsize{c}}}$. The top horizontal line is~$S_{\mbox{\scriptsize{c}}}$ for the RG---it is flat because 
the graph is the same regardless of~$p_{\mbox{\scriptsize{w}}}$. The three lines below correspond to the WS-I topologies. 
 For each value of~$p_{\mbox{\scriptsize{w}}}$, we generate~$150$ WS-I graphs.
Of the WS-I three lines, the top line corresponds, at each~$p_{\mbox{\scriptsize{w}}}$, to the topologies (among the 150 generated)   with maximum convergence rate, the medium line to the  average  convergence rate (averaged over the 150 random topologies generated), and the bottom line to the topologies (among the 150 generated) with worst convergence rate.  Similarly, the center and right panels on Fig.~\ref{Sc_WSI_6038} compare the eigenratio parameters~$\gamma$ (center panel) and the algebraic connectivity~$\lambda_{2}$ (right panel). For example, the RG improves by 50~\% the $\gamma$ eigenratio over the best WS-I topology (in this case for~$p_{\mbox{\scriptsize{w}}}=.8$.) 

\begin{figure}
\begin{center}
\includegraphics[height=2.1in, width=2.1in ] {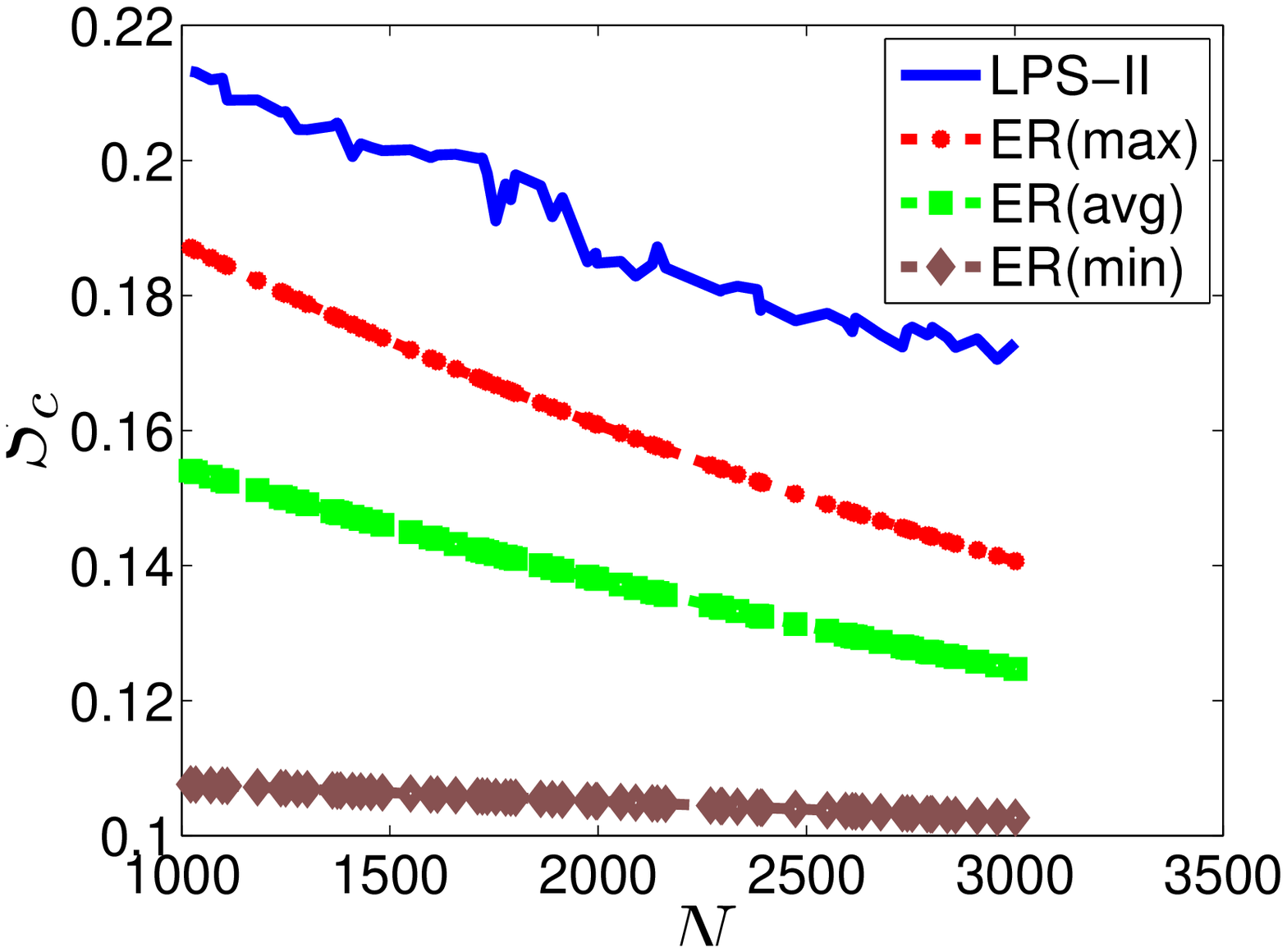}
\includegraphics[height=2.1in, width=2.1in ] {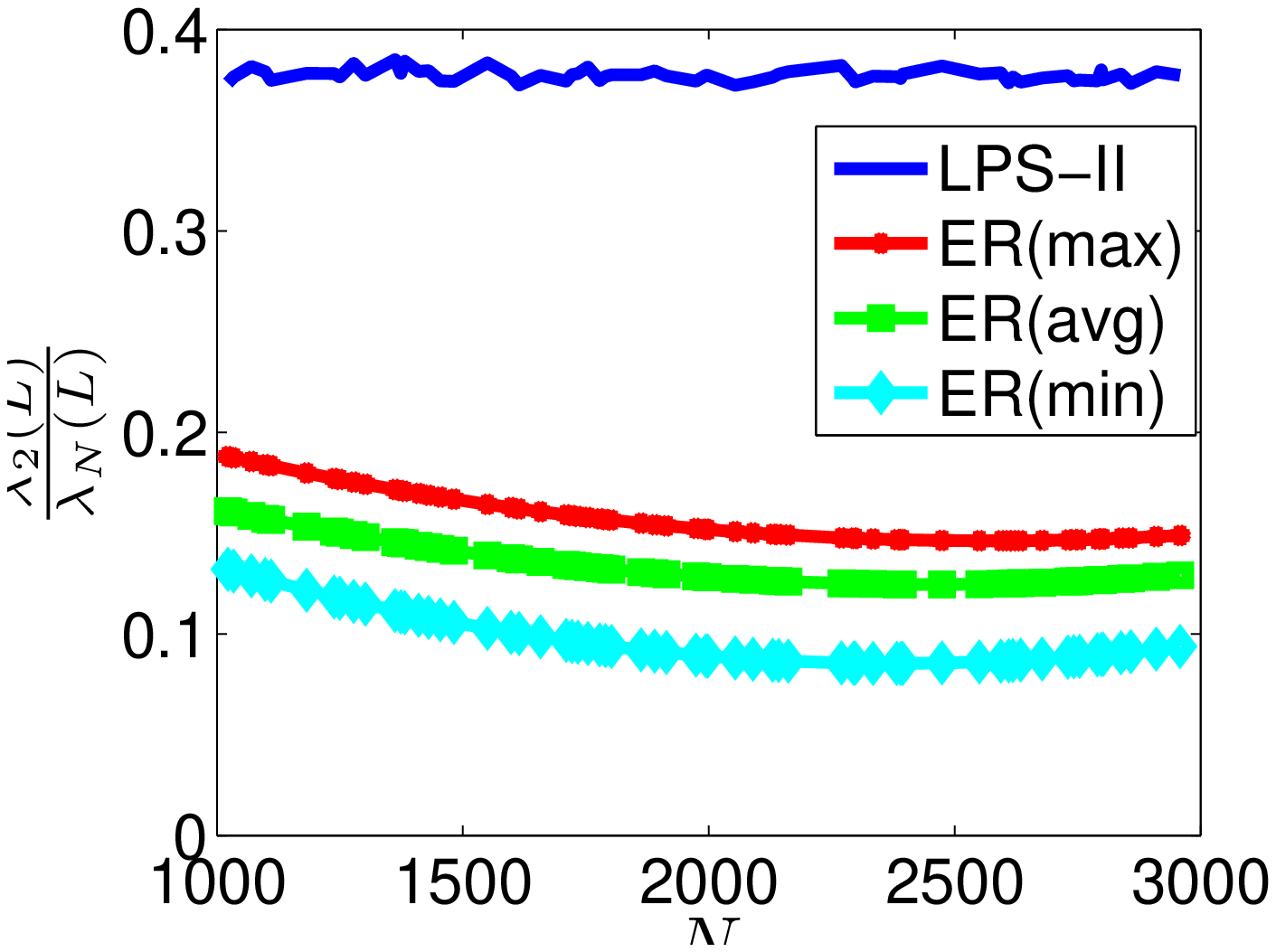}
\includegraphics[height=2.1in, width=2.1in ] {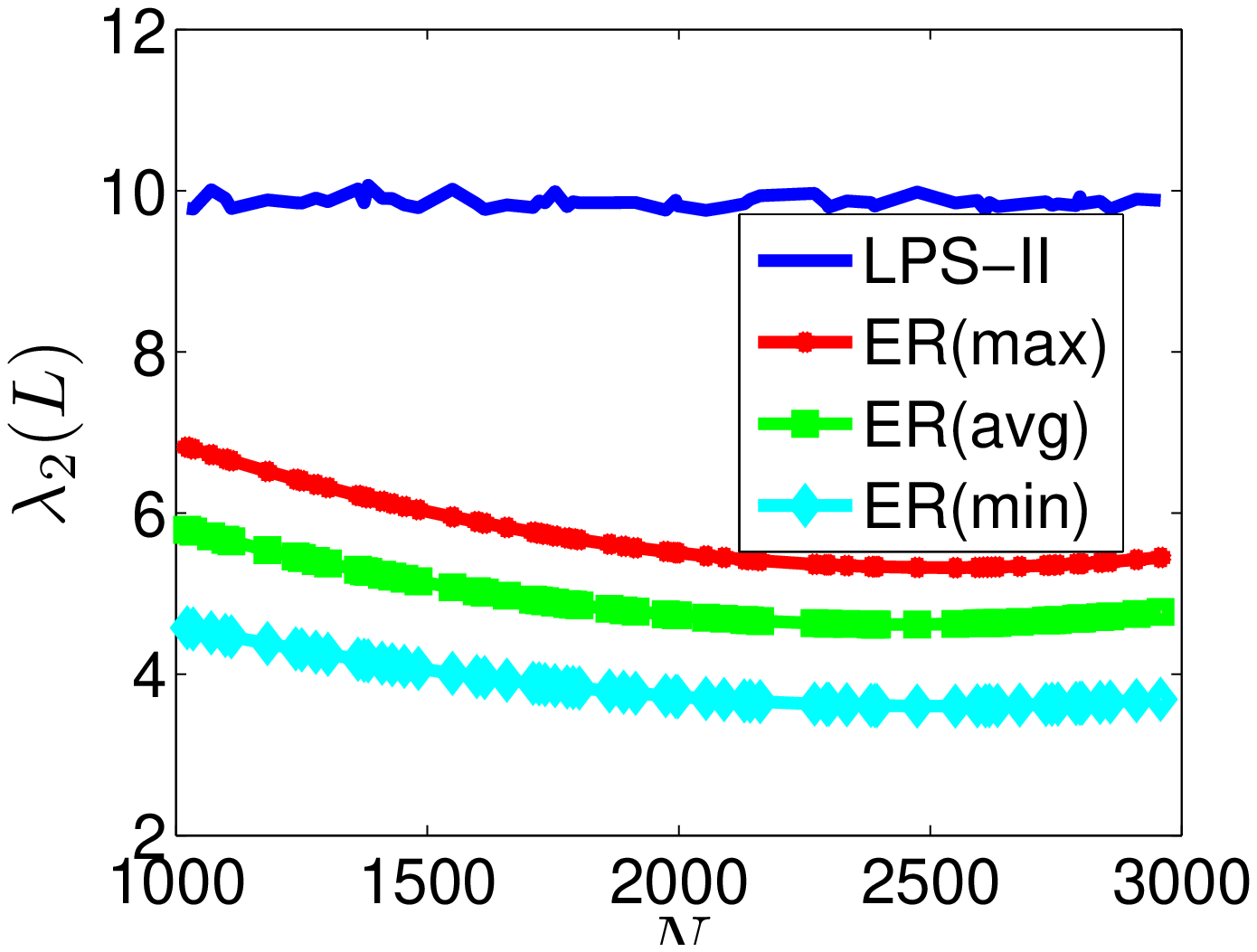}
\caption{Spectral properties of LPS-II and ER graphs, $k=18$, varying~$N$:
Left: Convergence speed $S_{\mbox{\scriptsize{c}}}$;
Center: eigenratio
$\gamma=\frac{\lambda_{2}(L)}{\lambda_{N}(L)}$; Right: algebraic connectivity
$\lambda_{2}$.} 
\label{eig_ER_ratio-2}
\label{eig_ER_ratio}
\label{eig_ER_lambda_2}\label{Sc_ER}
\end{center}
\end{figure}

{\bf Ramanujan graphs and Erd\"os-Ren\'{y}i graphs.}
We conclude this section
by comparing the LPS-II graphs with
the Erd\"{o}s-Ren\'{y}i graphs in Figs.~\ref{eig_ER_ratio} and~\ref{eig_ER_ratio-2}. Fig.~\ref{eig_ER_ratio} shows the results for topologies with different number of sensors~$N$ (plotted in the horizontal axis.) For each value of~$N$, we generated 200 random Erd\"os-Ren\'{y}i graphs. In the panels of both Figures, the top line illustrates the results for the RG, while the three lines below show the results for the Erd\"os-Ren\'{y}i graphs---among these three, the top line is the topology with best convergence rate among the 200~ER topologies, the middle plot is the averaged convergence rate, averaged over the 200 topologies, and the bottom line corresponds to the worst topologies. Again, for example, the~$\gamma$ parameter of the RG is about twice as large than the $\gamma$ parameter for the ER.

\section{Random Regular Ramanujan-Like Graphs}
\label{randomgraphscloseramanujan}
Section~\ref{ramanujanconstruction} explains the construction of the Ramanujan graphs. These graphs can be constructed only for certain values of~$N$, which may limit their application in certain practical scenarios. We describe here briefly biased random graphs that can be constructed with arbitrary number of nodes~$N$ and average degree, and whose performance closely matches that of Ramanujan graphs.  Reference~\cite{HetDegCorrPhy} argues that, in general, heterogeneity in the degree distribution reduces the eigenratio
$\gamma=\frac{\lambda_{2}(L)}{\lambda_{N}(L)}$. Hence, we try to construct
graphs that are regular in terms of the degree. There exist
constructions of random regular graphs, but these are difficult
to implement especially for very large number of vertices, see, e.g., 
\cite{Reg1,Reg2,Reg3,Reg4}, which are good
references on the construction and application of random regular
graphs. 

Ours is a  procedure that is simple to implement and constructs
random regular graphs, which we refer to as Random Regular
Ramanujan-Like~(R3L) graphs. 
Suppose, we want to construct a random regular graph with $N$
vertices and degree~$k$. Our construction starts from a regular graph of
degree~$k$, which we call the seed. The seed can be any regular
graph of degree~$k$, for example, the regular ring lattice with
degree~$k$ (see Section~\ref{results}.) We start by randomly choosing
(uniformly) a vertex (call it $v_{1}$.) In the next step, we
randomly choose a neighbor of $v_{1}$ (call it $v_{2}$), and we
also randomly choose a vertex not adjacent to $v_{1}$ (call it
$v_{3}$.) We now choose a neighbor of $v_{3}$ (call it $v_{4}$).
The next step consists of removing the edges between $v_{1}$ and
$v_{2}$, and between $v_{3}$ and $v_{4}$. Finally we add edges
between $v_{1}$ and $v_{3}$ and between $v_{2}$ and $v_{4}$. (Care
is taken so that no conflict arises in the process of removing and
forming the edges.) It is quite clear that after this sequence of
steps, the degree of each vertex remains the same and hence the
resulting graph remains $k$-regular. We repeat this sequence of
steps a sufficiently large number of times, which makes the
resulting graph to become random. Thus, staring with any $k$-regular graph,
we get a random regular graph with degree $k$. 

We now present numerical studies of the R3L graphs, which
show that these graphs have convergence properties that are very close to those of LPS-II graphs. Specifically, we focus on the eigenratio parameter $\gamma=\frac{\lambda_{2}(L)}{\lambda_{N}(L)}$.
\begin{figure}
\begin{center}
\includegraphics[height=2in, width=2in ]{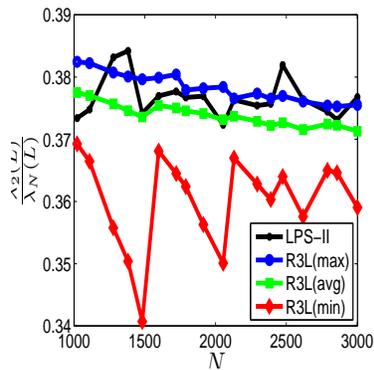}
\caption{LPS-II and R3L graphs, $k=18$, varying~$N$: Eigenratio $\gamma=\frac{\lambda_{2}(L)}{\lambda_{N}(L)}$.}
\label{lambda_ratio_R3L}
\end{center}
\end{figure}

Fig.~\ref{lambda_ratio_R3L} plots the eigenratio
$\gamma=\frac{\lambda_{2}(L)}{\lambda_{N}(L)}$ for the RG and the R3L graphs for varying number of nodes~$N$ and degree
$k=18$. We generate 100 R3L graphs  for each value of $N$. The top three lines correspond to the RG, the best R3L topologies, and the average value of~$\gamma$ over the 100~R3L graphs. We observe that the maximum values of
$\gamma=\frac{\lambda_{2}(L)}{\lambda_{N}(L)}$ are sometimes higher than those
obtained with the  LPS-II graphs. Note also that, on  average, the R3L graphs are quite close to the LPS-II graphs in terms
of the $\gamma=\frac{\lambda_{2}(L)}{\lambda_{N}(L)}$ ratio, even for large
values of~$N$. This study shows that the R3L graphs are a good alternative to the
LPS-II graphs with the added advantage that they can be generated for arbitrary number of nodes
$N$ and degree $k$.

\section{Conclusion}
\label{conclusion}
The paper studies the impact of network topology on the
convergence speed of distributed inference and  average-consensus. We
 derive that the convergence speed is governed by a graph spectral parameter, the
eigenratio $\gamma=\lambda_{2}(L)/\lambda_{N}(L)$ of the second largest and the largest eigenvalues of the graph Laplacian. We show that the class of non-bipartite Ramanujan graphs is essentially optimal. Numerical simulations verify the Ramanujan LPS-II graphs outperform the highly structured graphs, the Erd\"{o}s-Ren\'{y}i random graphs, and
graphs exhibiting the small-world property. We considered average-consensus and distributed detection with noiseless and noisy links. We derived for the distributed inference problem
 an analytical upper bound on the likelihood variance. For noiseless links, this bound shows that the local likelihood variances (and hence the
local probability of errors) converge to the global
likelihood variance (global probability of error) at a rate
determined by $\gamma$. With noisy links, we demonstrate that there is a maximum, optimal number of iterations before declaring a decision.  Finally, we introduced a novel biased construction of
random regular graphs (R3L graphs) and showed by numerical results
that their convergence performance tracks very closely that of the Ramanujan LPS-II graphs. R3L graphs address a main limitation of Ramanujan graphs that can be constructed only for very restricted number of nodes. In contrast, R3L graphs are simple to construct and can have an arbitrary number of nodes~$N$ and degree~$k$.

\addcontentsline{toc}{section}{{\LARGE \mathbf{Appendix}}}
\appendix
\label{appendix1}
{\begin{definition}[Group]: A group $X$ is  a non-empty collection
of elements, with a binary operation ``.'' defined on them, such
that the following properties are satisfied:
\begin{enumerate}
\item If $a,b\in X$, then $a.b\in X$ (closure property.) \item If
$a,b,c\in X$, then $a.(b.c)=(a.b).c$ (associative property.) \item
There exists an element $e\in X$, such that for any element $a\in
X$, $a.e=e.a=a$ (identity element.) 
\item  $\forall a\in X$,  there
exists  $a^{-1}\in X$,  the inverse of $a$, such
that $a.a^{-1}=a^{-1}.a=e$ (inverse.)
\end{enumerate}
The group $X$ is called abelian if the ``.'' operation is
commutative, that is, for any $a,b\in X$, $a.b=b.a$.
\end{definition}
\begin{definition}[Field]: A field $F$ is  a non-empty collection
of elements, with the following properties: \\ There exists a
binary operation ``+'' on the elements of $F$ such that,
\begin{enumerate}
\item If $a,b\in F$, then $a+b\in F$. \item If $a,b\in F$, then
$a+b=b+a$. \item If $a,b,c\in F$, then $a+(b+c)=(a+b)+c$. \item
There exists an element 0 (zero) $\in F$, such that for any
element $a\in F$, $a+0=a$. \item If $a\in F$, then there exists an
element $(-a)\in F$, such that $a+(-a)=(-a)+a=0$.
\end{enumerate}
There exists another binary operation ``.'' on the elements of $F$
such that,
\begin{enumerate}
\item If $a,b\in F$, then $a.b\in F$. \item If $a,b\in F$, then
$a.b=b.a$. \item If $a,b,c\in F$, then $a.(b.c)=(a.b).c$. \item
There exists a non-zero element 1 (one) $\in F$, such that for any
element $a\in F$, $a.1=a$. \item For every non-zero element $a\in
F$, there exists an element $a^{-1}\in F$, such that $a.a^{-1}=1$.
\item If $a,b,c\in F$, then $a.(b+c)=a.b+a.c$.
\end{enumerate}
\end{definition}

\emph{Congruence.} For integers $a,b,c$, the statement $a$ is
congruent to $b$ modulo $c$, or $a\equiv b\bmod(c)$ implies that
$(a-b)$ is divisible by $c$.

\emph{Quadratic Residue.} For integers $a,b$, the statement $a$ is a
quadratic residue modulo $b$ implies that there exists an integer
$c$ such that $c^{2}\equiv a\bmod(b)$.

\begin{definition}[Legendre Symbol]: For an integer $a$ and a prime $p$, the Legendre
symbol $\left(\frac{a}{p}\right)$ is
\begin{equation}
\left(\frac{a}{p}\right) = \left\{ \begin{array}{rl}
                                    0 & \mbox{if $p$ divides $a$}
                                    \\
                                    1 & \mbox{if $a$ is a
                                    quadratic residue modulo $p$}
                                    \\
                                    -1 & \mbox{if $a$ is a
                                    quadratic non-residue modulo
                                    $p$}
                                    \end{array}
                                    \right.
\end{equation}
\end{definition}

\emph{PSL(2,$Z/qZ$).} For a prime $q$, the set $Z/qZ=\{0,1,..,q-1\}$
is the field of integers modulo $q$. To define the group
PSL($2,Z/qZ$) (Projective Special Linear Group), first
consider the set of $2\times 2$ matrices over the field $Z/qZ$,
whose determinants are non-zero quadratic residues modulo $q$.
Next, define an equivalence relation on this set, such that two
matrices are in the same equivalence class, if one is a non-zero
scalar multiple of the other. The PSL($2,Z/qZ$) group is then the
set of all these equivalence classes. Think of each
element of PSL($2,Z/qZ$) as a $2\times 2$ matrix over the field
$Z/qZ$, whose determinant is a non-zero quadratic residue modulo
$q$, and whose second row can be represented as either (0,1) or
(1,$a$), where $a$ being any element of $Z/qZ$, \cite{SiamMath}.
The $p+1$ generators discussed in the paper, belong to the
PSL(2,$Z/qZ$) group, because their determinants are $p\bmod(q)$
and by assumption, $p$ is a quadratic residue modulo $q$ or
$\left(\frac{p}{q}\right)=1$ for the non-bipartite Ramanujan
graphs we use in this paper.

\emph{Linear Fractional Transformation.} Let $P^{1}(F_{q}) =
\{0,1,...,q-1,\infty\}$ and $\left(
\begin{array}{ll} a & b \\ c & d \end{array}\right) $ be a $2\times 2$
matrix. Then a linear fractional transformation on $P^{1}(F_{q})$
is defined by the mapping,
\begin{equation}
x\longmapsto \frac{ax+b}{cx+d}\bmod(q)
\end{equation}
for every element $x\in P^{1}(F_{q})$, with the usual assumptions
that $\frac{z}{0} = \infty$ for $z\neq 0$, and $\frac{a\infty +
b}{c\infty + d}=\frac{a}{c}$.

\begin{definition}[Bipartite graph]: A bipartite graph is a graph
in which the vertex set can be partitioned into two disjoint
subsets, such that no two vertices in the same subset are
adjacent.
\end{definition}
}

\bibliographystyle{IEEEtran}
\bibliography{IEEEabrv,BibOptEqWeights}

\begin{thebibliography}{10}
\providecommand{\url}[1]{#1}
\csname url@rmstyle\endcsname
\providecommand{\newblock}{\relax}
\providecommand{\bibinfo}[2]{#2}
\providecommand\BIBentrySTDinterwordspacing{\spaceskip=0pt\relax}
\providecommand\BIBentryALTinterwordstretchfactor{4}
\providecommand\BIBentryALTinterwordspacing{\spaceskip=\fontdimen2\font plus
\BIBentryALTinterwordstretchfactor\fontdimen3\font minus
  \fontdimen4\font\relax}
\providecommand\BIBforeignlanguage[2]{{%
\expandafter\ifx\csname l@#1\endcsname\relax
\typeout{** WARNING: IEEEtran.bst: No hyphenation pattern has been}%
\typeout{** loaded for the language `#1'. Using the pattern for}%
\typeout{** the default language instead.}%
\else
\language=\csname l@#1\endcsname
\fi
#2}}

\bibitem{SensNets:Tenney81}
R.~R. Tenney and N.~R. Sandell, ``Detection with distributed sensors,''
  \emph{{IEEE} Trans. Aerosp. Electron. Syst.}, vol. {AES}-17, pp. 98--101,
  July 1981.

\bibitem{SensNets:Tsitsiklis88}
J.~N. Tsitsiklis, ``Decentralized detection by a large number of sensors,''
  \emph{MCSS}, vol.~1, no.~2, pp. 167--182, 1988.

\bibitem{SensNets:Tsitsiklis93}
------, ``Decentralized detection,'' \emph{In "Advances in Statistical Signal
  Processing: Vol~2~-~Signal Detection," \emph{H. V. Poor, and John B. Thomas,
  eds., JAI Press, Greenwich, CT,}}, pp. 297--344, Nov. 1993.

\bibitem{SensNets:Willett00}
P.~K. Willett, P.~F. Swaszek, and R.~S. Blum, ``The good, bad and ugly:
  distributed detection of a known signal in dependent gaussian noise,''
  \emph{{IEEE} Trans. Signal Processing}, vol.~48, p. 3266–3279, Dec. 2000.

\bibitem{SensNets:Varshney}
P.~K. Varshney, \emph{Distributed Detection and Data Fusion}.\hskip 1em plus
  0.5em minus 0.4em\relax New York: Springer-Verlag, 1996.

\bibitem{SensNets:Blum97}
R.~S. Blum, S.~A. Kassam, and H.~V. Poor, ``Distributed detection with multiple
  sensors: {P}art {II}--{A}dvanced topics,'' \emph{Proc. {IEEE}}, vol.~85, pp.
  64--79, Jan. 1997.

\bibitem{SensNets:Chamberland03}
J.-F. Chamberland and V.~V. Veeravalli, ``Decentralized detection in sensor
  networks,'' \emph{{IEEE} Trans. Signal Processing}, vol.~51, pp. 407--416,
  Feb. 2003.

\bibitem{SensNets:AldosariAsilomar05}
S.~A. Aldosari and J.~M.~F. Moura, ``Distributed detection in sensor networks:
  connectivity graph and small-world networks,'' in \emph{Asilomar Conference
  on Signals, Systems, and Computers}, 2005.

\bibitem{aldosarimouramay06}
------, ``Topology of sensor networks in distributed detection,'' in
  \emph{ICASSP'06, IEEE International Conference on Signal Processing}, May
  2006.

\bibitem{erdosrenyi59}
P.~Erd{\"o}s and A.~R{\'e}nyi, ``On random graphs,''
  \emph{Publ.~Math.~Debrecen}, vol.~6, pp. 290--291, 1959.

\bibitem{erdosrenyi60}
------, ``On the evolution of random graphs,''
  \emph{Publ.~Math.~Inst.~Hung.~Acad.~Sciences (Magyar
  Tud.~Akad.~Mat.~Kutat\'{o} Int.~K\"{o}zl.)}, vol.~5, pp. 17--61, 1960.

\bibitem{erdosrenyi61}
------, ``On the evolution of random graphs,''
  \emph{Bull.~Inst.~Internat.~Statist.}, vol.~38, pp. 343--347, 1961.

\bibitem{gilbert59}
E.~N. Gilbert, ``Random graphs,'' \emph{Annals of Mathematical Statistics},
  vol.~30, pp. 1141--1144, 1959.

\bibitem{SensNets:Bollobas98}
B.~Bollob\'as, \emph{Modern Graph Theory}.\hskip 1em plus 0.5em minus
  0.4em\relax New York, NY: Springer Verlag, 1998.

\bibitem{SensNets:Watts98}
D.~J. Watts and S.~H. Strogatz, ``Collective dynamics of small-world
  networks,'' \emph{Nature}, vol. 393, pp. 440--442, 1998.

\bibitem{SensNets:KleinbergNature2000}
J.~M. Kleinberg, ``Navigation in a small world,'' \emph{Nature}, vol. 406, p.
  845, Aug. 2000.

\bibitem{SensNets:KleinbergThComp2000}
------, ``The small-world phenomenon: An algorithmic perspective,'' in
  \emph{Proc. of the thirty-second annual ACM symposium on Theory of
  computing}, vol.~2, Portland, Oregon, 2000, pp. 63--170.

\bibitem{SensNets:Olfati04}
R.~Olfati-Saber and R.~M. Murray, ``Consensus problems in networks of agents
  with switching topology and time-delays,'' \emph{IEEE Trans. Automat.
  Contr.}, vol.~49, pp. 1520--1533, 2004.

\bibitem{SensNets:Fiedler73}
M.~Fiedler, ``Algebraic connectivity of graphs,'' \emph{{C}zechoslovak.
  {M}athematical {J}ournal}, vol.~23, no.~98, pp. 298--305, 1973.

\bibitem{FanChung}
F.~R.~K. Chung, \emph{Spectral {G}raph {T}heory}.\hskip 1em plus 0.5em minus
  0.4em\relax American Mathematical Society, 1997.

\bibitem{SensNets:Xiao04}
L.~Xiao and S.~Boyd, ``Fast linear iteration for distributed averaging,''
  \emph{Syst. Contr. Lett.}, vol.~53, pp. 65--78, Sept. 2004.

\bibitem{LPS}
A.~Lubotzky, R.~Phillips, and P.~Sarnak, ``Ramanujan graphs,''
  \emph{Combinatorica}, vol.~8, no.~3, pp. 261--277, 1988.

\bibitem{Alon}
N.~Alon, ``Eigenvalues and expanders,'' \emph{Combinatorica}, vol.~6, pp.
  83--96, 1986.

\bibitem{Murty}
M.~R. Murty, ``Ramanujan graphs,'' \emph{J. Ramanujan Math. Soc}, vol.~18,
  no.~1, pp. 1--20, 2003.

\bibitem{Margulis}
G.~Margulis, ``Explicit group-theoretical constructions of combinatorial
  schemes and their application to the design of expanders and concentrators,''
  \emph{J. Probl. Inf. Transm.}, vol.~24, no.~1, pp. 39--46, 1988.

\bibitem{Morgenstern}
M.~Morgenstern, ``Existence and explicit construction of $q+1$ regular
  {R}amanujan graphs for every prime power $q$,'' \emph{J. Comb. Theory, ser.
  B}, vol.~62, pp. 44--62, 1994.

\bibitem{jacobi_proof}
G.~Andrews, S.~B. Ekhad, and D.~Zeilberger, ``A short proof of {J}acobi's
  formula for the number of representations of an integer as a sum of four
  squares,'' \emph{Amer. Math. Monthly}, vol. 100, pp. 273--276, 1993.

\bibitem{Lafferty}
A.~Berger and J.~Lafferty, ``Probabilistic decoding of low-density {C}ayley
  codes.''

\bibitem{VanTreesPartI}
H.~L.~V. Trees, \emph{Detection, Estimation, and Modulation Theory:
  Part~{I}}.\hskip 1em plus 0.5em minus 0.4em\relax New York, NY: John Wiley \&
  Sons, 1968.

\bibitem{HetDegCorrPhy}
A.~.~E. Motter, C.~Zhou, and J.~Kurths, ``Network synchronization, diffusion,
  and the paradox of heterogeneity,'' \emph{Physical Review E}, vol.~71, 2005.

\bibitem{Reg1}
E.~A. Bender and E.~R. Canfield, ``The asymptotic number of non-negative
  integer matrices with given row and column sums,'' \emph{Journal of
  Combinatorial Theory, Series A}, vol.~24, pp. 296--307, 1978.

\bibitem{Reg2}
B.~Bollob\'{a}s, ``A probabilistic proof of an asymptotic formula for the
  number of labelled regular graphs,'' \emph{European Journal of
  Combinatorics}, vol.~1, pp. 311--316, 1980.

\bibitem{Reg3}
N.~C. Wormald, ``Generating random regular graphs,'' \emph{Journal of
  Algorithms}, vol.~5, pp. 247--280, 1984.

\bibitem{Reg4}
------, ``Models of random regular graphs,'' \emph{London Mathematical Society
  Lecture Note Series}, vol. 267, pp. 239--298, 1999.

\bibitem{SiamMath}
Y.~Kohayakawa, V.~Rodl, and L.~Thoma, ``An optimal algorithm for checking
  regularity,'' \emph{SIAM J. on Computing}, vol.~32, no.~5, pp. 1210--1235,
  2003.

\end{thebibliography}

\end{document}